# Topotactic Reduction-Driven Crystal Field Excitations in Brownmillerite Manganite Thin Films

Feng Jin, Shiyu Fan,* Mingqiang Gu, Qiming Lv, Min Ge, Zixun Zhang, Jinfeng Zhang, Jingdi Lu, Taehun Kim, Vivek Bhartiya, Zhen Huang, Lingfei Wang, Valentina Bisogni, Jonathan Pelliciari,* and Wenbin Wu*

Feng Jin, Qiming Lv, Min Ge, Zixun Zhang, Jinfeng Zhang, Jingdi Lu, Lingfei Wang, Wenbin Wu
Hefei National Research Center for Physical Sciences at the Microscale
University of Science and Technology of China
Hefei 230026, China
E-mail: wuwb@ustc.edu.cn

Shiyu Fan, Taehun Kim, Vivek Bhartiya, Valentina Bisogni, Jonathan Pelliciari
National Synchrotron Light Source II
Brookhaven National Laboratory
Upton, NY 11973, USA
E-mail: sfan1@bnl.gov; pelliciari@bnl.gov

Mingqiang Gu
Department of Physics
Southern University of Science and Technology
Shenzhen 518055, China

Zhen Huang, Wenbin Wu
Institutes of Physical Science and Information Technology
Anhui University
Hefei 230601, China

Zhen Huang,
Information Materials and Intelligent Sensing Laboratory of Anhui Province
Anhui University
Hefei 230601, China

Wenbin Wu
High Magnetic Field Laboratory
Chinese Academy of Sciences
Hefei 230031, China

These authors contributed equally: Feng Jin, Shiyu Fan, and Mingqiang Gu.

**ABSTRACT：**

Topotactic reduction of perovskite oxides offers a powerful approach for discovering novel phenomena, such as superconducting infinite-layer nickelates and polar metallicity, and is commonly accompanied by the emergence of multiple valence states and/or complex crystal fields of transition metals. However, understanding the complex interplay between crystal chemistry, electronic structure, and physical properties at the spin- and orbital-resolved levels in these reduced systems remains elusive. Here, we combine x-ray absorption spectroscopy, resonant inelastic x-ray scattering (RIXS), and density functional theory calculations to uncover topotactic metal-insulator transition and orbital-specific crystal field excitations in brownmillerite $La_{0.67}Ca_{0.33}MnO_{2.5}$ thin films. We reveal the Mn valence states to be predominantly $Mn^{2+}/Mn^{3+}$, along with their corresponding populations at octahedral and tetrahedral sites, which effectively weaken the Mn-O hybridization compared to the parent perovskite phase. As a result, $La_{0.67}Ca_{0.33}MnO_{2.5}$ films exhibit an antiferromagnetic insulating ground state. Moreover, by combining the RIXS measurements on selected single-valence manganites, specifically MnO, $LaMnO_3$, and $CaMnO_3$, with orbital- and spin-resolved density-of-states calculations, we identify the *dd* excitations of octahedrally and tetrahedrally coordinated $Mn^{2+}/Mn^{3+}$ ions, directly linking the microscopic electronic structure to the macroscopic magnetic/electrical properties.

## 1. Introduction

Topotactic reduction, involving the removal of oxygen from low concentrations up to tens of atomic percent, facilitates transitions from perovskite ($ABO_3$) to brownmillerite ($ABO_{2.5}$) and infinite-layer ($ABO_2$) phases.[1-8] As a result, transition metal oxides may host multiple crystallographic sites and oxidation states at the B site, enabling diverse functionalities.[1–4] The combination of epitaxial growth and topotactic reduction[3,9-15] has even led to the discovery of superconductivity in $d^9$ infinite-layer nickelates. Nevertheless, synthesizing high-quality, stoichiometrically precise reduced films remains challenging due to the need for accurate control over oxygen content while minimizing secondary phases, crystal defects, and excess/deficient oxygen vacancies.[1–3,16]

Unlike infinite-layer or perovskite structures, brownmillerite oxides feature alternating octahedral ($O_h$) and tetrahedral ($T_d$) layers, making them suitable for ionic conducting and anion insertion applications.[1,2,5,10,11,18,19,27] Such unique structure is prevalent in a plethora of materials, including $AMnO_{2.5}$,[19-26] (Ca,Sr)$FeO_{2.5}$,[11,12,14] and (La,Ca,Sr)$CoO_{2.5}$,[5,9,10,13,15,17,18] to name a few. Interestingly, these materials exhibit emergent phenomena (such as magnetism) and are relevant to ferroelectric devices[19] or magnetic heterostructures[13]. However, understanding the relationship between their electronic structures and macroscopic properties remains elusive due to complex crystal fields and/or multiple valence states of the B-site atoms. Furthermore, the strong coupling between multiple degrees of freedom and the relevance of factors such as electron hopping amplitude, on-site Coulomb repulsion, and Hund's coupling, all in the several-eV range, add further complexity.[17-19]

In brownmillerite manganites, the reduction of perovskite manganites induces ordered oxygen vacancies, disrupting super- and double-exchange interactions between manganese atoms and altering their oxidation states.[20-25] The resulting balance of Mn ions with different oxidation states and electronic configurations (including distinct orbital and spin components), offers an opportunity to fine-tune functionalities like electrical transport and magnetism, as demonstrated in various magnetic materials[15,17,26-31] and superconducting nickelates.[32,33] However, a detailed microscopic understanding of these properties requires an experimental tool that can probe all electronic degrees of freedom while distinguishing oxidation states and local crystal fields. Resonant inelastic x-ray scattering (RIXS), a photon-in photon-out scattering technique, is uniquely capable of probing spin, charge, orbital, and lattice excitations via its resonant process.[32,34-38] This resonance allows for selective detection of oxidation states and local coordination environments. Compared to previous Mn *K* edge RIXS on perovskite manganites, Mn *L*-edge measurements offer higher energy resolution and more directly probe Mn 3*d* multiplet excitations.[39–41] Unfortunately, this important aspect has been underutilized in exploring the electronic structure of materials with multiple oxidation states and complex coordination environments.[39-43]

Here, we demonstrate the synthesis of coherently epitaxial $La_{0.67}Ca_{0.33}MnO_{2.5}$ ($LCMO_{2.5}$) thin films with a brownmillerite structure, achieved through a topotactic phase transformation from perovskite $La_{0.67}Ca_{0.33}MnO_3$ ($LCMO_3$). The ground state transitions from a half-metallic ferromagnetism in $LCMO_3$ to an insulating antiferromagnetic phase in $LCMO_{2.5}$. Density functional theory (DFT) calculations indicate that $LCMO_{2.5}$ adopts the *Pbnm* space group with a G-type antiferromagnetic structure, featuring ordered chain arrangements in the tetrahedral layers with alternating orientations in adjacent layers. Using x-ray absorption (XAS), RIXS, and DFT calculations, we further identify the electronic configurations of the Mn ions: $Mn^{2+}$ ions occupy all the tetrahedral sites and one-third of the octahedral sites, while $Mn^{3+}$ ions occupy the remaining octahedral sites. By comparing the RIXS spectra of single-valence systems (MnO, $LaMnO_3$, and $CaMnO_3$) with those of $LCMO_{2.5}$, as well as the joint density of states (DOS), we disentangle the orbital excitations arising from the multiple Mn oxidation states. Our results provide a comprehensive understanding of the energy scales of competing interactions, including Jahn-Teller splitting, Hund's coupling, and intersite electron hopping. This approach provides valuable microscopic insights into the significant changes in the transport properties upon chemical reduction.

## 2. Results and Discussion

### A. Structural characterization of the brownmillerite $LCMO_{2.5}$ film

The brownmillerite structure of $LCMO_{2.5}$ is evidenced by x-ray diffraction (XRD) $2\theta-\omega$ linear scans in **Figure 1**. The appearance of new Bragg peaks indicates the change in structure from a perovskite to a brownmillerite. As schematically shown in Figure 1a, the brownmillerite structure consists of alternately stacked octahedral ($MnO_6$) and tetrahedral ($MnO_4$) layers, doubling the *c* lattice constant. At the same time, the crystalline periodicity along the *a* and *b* directions remains unchanged (Table S1, Supporting Information, and Figure S1, Supporting Information). Compared with the reflections of $LCMO_3$ films, an additional set of reflections appears in Figure 1b, confirming the doubling of the *c* lattice constant. Due to the volume expansion caused by ordered oxygen vacancies, the angle of the (008) brownmillerite reflection shifts to a lower value of 44.457°, compared to the angle of the (004) perovskite reflection. The diffraction angle of the (008) reflection agrees well with the reported value for $La_{0.75}Ca_{0.25}MnO_{2.5}$ from powder diffraction.[29,44] Accordingly, we calculate the *c* lattice constant to be 16.133 Å. Moreover, Laue fringes appear around the (008) reflection of $LCMO_{2.5}$, and the full width at half maximum (FWHM) of its rocking curve is 0.0567° (Figure S2, Supporting Information), indicating the high quality of $LCMO_{2.5}$. According to these Laue fringes, we calculated the film thickness, which is consistent with the predetermined 24 nm.

To further scrutinize the heteroepitaxial growth and strain states, high-resolution reciprocal lattice mappings around the (206) and (2 0 12) reflections from $LCMO_3$ and $LCMO_{2.5}$ films were performed. As shown in Figure 1c, the elongation of these reflections is attributed to the confined finite size along the film normal direction. The in-plane lattice parameters of both $LCMO_3$ and $LCMO_{2.5}$ films are consistent with those of $NdGaO_3$(001) [NGO(001)] substrates, implying coherently epitaxial growth. This enables the calculation of the unit-cell volume as approximately 482.39 $Å^3$. As a result, the $LCMO_{2.5}$ film exhibits a +0.24% tensile strain along the [010] direction and a -3.95% compressive strain along the [100] direction (Figure S1, Supporting Information). Despite the significantly large compressive strain, $LCMO_{2.5}$ film maintains a coherent state, possibly due to the distortion and flexible geometrical folding of $MnO_6$ octahedra and $MnO_4$ tetrahedra along the [100] direction.

To directly visualize the alternately stacked octahedral and tetrahedral layers along the growth direction, we also performed high-angle annular dark-field (HAADF) scanning transmission electron microscopy (STEM) images of $LCMO_{2.5}$ along the [010] zone axis of NGO. The removal of oxygen enhances the Coulomb repulsion between cations,[24] inducing the volume expansion. Consequently, the tetrahedral layers have a larger lattice spacing ($d_{Tetra}$) along the out-of-plane direction compared to the smaller lattice spacing ($d_{Octa}$) of octahedral layers. This removal of oxygen ions and the re-arrangement of Mn sites during reduction are evidenced by the periodic variation in interplanar spacing, as shown in the left panel of Figure 1d. Obviously, we can identify the alternating larger $d_{Tetra}$ and smaller $d_{Octa}$ spacings. In addition, the HAADF

STEM image in the right panel of Figure 1d clearly shows a sharp interface between $LCMO_{2.5}$ and NGO.

## B. Determining the electronic ground state of $LCMO_{2.5}$

The perovskite $LCMO_3$ film exhibits a half-metallic feature, whereas the brownmillerite $LCMO_{2.5}$ film possesses an insulating and nonmagnetic behavior. In $LCMO_3$, the double-exchange interaction between manganese ions bridged by oxygen plays an essential role in allowing $e_g$ electrons to hop from one site to another,[45,46] inducing itinerant ferromagnetism due to strong Hund's coupling, as shown in **Figure 2**a,b. In contrast, for $LCMO_{2.5}$, the alternating octahedral and tetrahedral layers, coupled with lower Mn chemical valence states, result in an insulating behavior and a non-ferromagnetic state. Notably, due to the in-plane anisotropic structure of unique tetrahedral chains along the [010] direction in $LCMO_{2.5}$ films (**Figure 3**a), anisotropic electronic transport properties appear, as shown in Figure 2c.

Next, we performed DFT calculations to understand experimentally measured structural and electronic properties. Structurally, there are three possible arrangements for the tetrahedral chains in brownmillerite-type frameworks, corresponding to the *Pbnm*, *Im2c*, and *Pmca* space groups, as shown in Figure 3a and Figure S3 (Supporting Information).[47] Using a G-type antiferromagnetic structure in our DFT calculations, we found that the $LCMO_{2.5}$ film adopts a *Pbnm* structure, which is always the most stable in the low-strain states (Figures S4 and S5, Supporting Information).

Considering the three positions of oxygen ions and the two coordination environments of manganese ions, we further divided the DOS into contributions from tetrahedral and octahedral manganese ions, and intermediate, tetrahedral, and octahedral oxygen ions. As reported in Figure 3b, the DOS of oxygen and manganese ions in the octahedral layers play a pivotal role near the Fermi level in $LCMO_{2.5}$. The valence band maximum and conduction band minimum are mainly contributed by the octahedral manganese $e_g$ states due to Jahn-Teller splitting, resulting in a band gap in $LCMO_{2.5}$, in sharp contrast to $LCMO_3$.[45] These results suggest that the topotactic structural transition is accompanied by a change in the ground state from a half-metallic ferromagnetism to an insulating antiferromagnetism. Moreover, the DFT-predicted electronic band structure reveals significant anisotropy in the effective electron masses between the [100] and [010] directions, as illustrated in Figure 3c. This result explains the observed anisotropic electrical transport properties in Figure 2c.

However, the significant role of Mn 3*d* energy levels in transitioning from octahedral to tetrahedral coordination needs further exploration.

### C. Identifying multiple Mn valence states in $LCMO_{2.5}$

We now investigate the Mn valence states and their associated crystal-field excitations in $LCMO_{2.5}$ using XAS at the Mn *L* and O *K* edges and RIXS at the Mn *L* edge. Compared to optical spectroscopy, the dipole and spin selection rules are not essential to RIXS which makes it a sensitive technique to probe localized *dd* excitations between the multiplet states of transition metal ions.[37,42] **Figure 4**a showcases the experimental geometry of the XAS and RIXS experiments. To enhance the inelastic features, we align the crystallographic $[111]_C$ direction (in the pseudo-cubic notation) with the scattering plane, as the elastic peak is suppressed away from specular reflection. Figure 4b displays a comparison of the XAS between $LCMO_{2.5}$ and $LCMO_3$ at the Mn-*L* and O-*K* edges. A sharp peak at 639.5 eV appears in $LCMO_{2.5}$, but absent in $LCMO_3$. This feature corresponds to the $Mn^{2+}$ valence state,[27] and its strong intensity indicates a significant amount of $Mn^{2+}$ after chemical reduction. In fact, our DFT calculation shows that all manganese ions in the $MnO_4$ tetrahedra have a $Mn^{2+}$ valence state, as shown in Figure S6 (Supporting Information), consistent with the experimental data.

According to DFT calculations, not only the tetrahedral sites, but some of the octahedral sites also exhibit the $Mn^{2+}$ valence state. This can be expected by examining the chemical formula and the valence states of the elements in $LCMO_{2.5}$. The integer chemical formula of $LCMO_{2.5}$ can be written as $La_4Ca_2Mn(octa)_3Mn(tetra)_3O_{15}$, representing six units of $LCMO_{2.5}$, where Mn(octa) and Mn(tetra) denote Mn atoms at octahedral and tetrahedral sites, respectively. By considering the balance of valence states as $La^{3+}$, $Ca^{2+}$, $Mn(tetra)^{2+}$, and $O^{2-}$, it follows that the three Mn(octa) ions must collectively lose 8 electrons. Given the 1/3-doping at the A-site, two possible electronic configurations for Mn(octa) sites arise: i) $2{\times}Mn^{3+}+1{\times}Mn^{2+}$, or ii) $2{\times}Mn^{2+}+1{\times}Mn^{4+}$. However, the configuration ii) is not feasible because $Mn^{2+}$ and $Mn^{4+}$ ions, with $d^5$ and $d^3$ configurations, respectively, would induce octahedral breathing distortions—expansion for $Mn^{2+}$ and contraction for $Mn^{4+}$-while simultaneously suppressing Jahn-Teller distortions. As a result, the 2:1 population ratio cannot support them to form an intact octahedral network (Figure S7, Supporting Information), making it geometrically inconvenient. Therefore, the configuration i) should be the realistic electronic configuration, which is confirmed by checking the charge states in DFT calculations. In this sense, two-thirds of the Mn sites in $LCMO_{2.5}$ exhibit an $Mn^{2+}$ character.

For the O-*K* edge XAS, the pre-edge peak between 530 – 533.5 eV is observed for both $LCMO_{2.5}$ and $LCMO_3$, which is associated with the Mn 3*d* and O 2*p* orbital hybridization.[43] Compared to $LCMO_3$, this peak shifts to slightly higher energy in $LCMO_{2.5}$, and its intensity is significantly suppressed, indicating a weaker orbital hybridization. As shown in Figure 3b, the O 2*p* orbital primarily hybridizes with the $Mn^{3+}$ 3*d* orbitals in the octahedral environment. In $LCMO_{2.5}$, the emergence of tetrahedral sites occupied by $Mn^{2+}$ ions reduces the concentration of $Mn^{3+}$, thereby suppressing the spectral weight of the pre-edge hybridization peak. The peaks above 535 eV are associated with the transitions from O 2*p* to other higher energy unoccupied bands, and these do not display significant differences between $LCMO_{2.5}$ and $LCMO_3$. In brief, the XAS at both the Mn-*L* and O-*K* edges demonstrate that $Mn^{2+}$ plays a dominant role in $LCMO_{2.5}$.

The multi-valence nature of the topotactically reduced $LCMO_{2.5}$ complicates the identification of the orbital configurations of different Mn sites. To address this issue, we employ RIXS at the Mn $L_3$ edge to probe the local nature of Mn ions. Compared to previous RIXS measurements at the Mn *K* edge,[39–41] our measurements at the Mn $L_3$ edge has higher energy resolution and more directly access the multiplet excitations of Mn 3*d* orbitals, as the 2*p* - 3*d* transition is dipole-allowed. **Figure 5**a depicts the RIXS intensity color map as a function of incident x-ray energy and energy loss for $LCMO_{2.5}$. The complexity of Mn valence states and a combination of tetrahedral and octahedral crystal environments around Mn atoms make the RIXS spectra rich in orbital excitations. To assign the peaks in XAS, we compare the experimental results with DFT calculations of the electronic DOS for qualitative insight. Since XAS is proportional to the unoccupied DOS, we overlay the XAS spectra of $LCMO_{2.5}$ with the calculated empty DOS of Mn 3*d* orbitals projected on both spin-up and spin-down channels above the Fermi level. Strikingly, the XAS lineshape is well captured by the DFT calculations. The peaks at ≈ 640 and 641 eV in XAS have a mixed character of $Mn^{2+}$ in the tetrahedral site and $Mn^{3+}$ in the octahedral site, while the peak centered at 643.5 eV corresponds to the $Mn^{2+}$ $e_g$ bands in the octahedral site. As we will discuss, this assignment agrees well with the RIXS intensity map.

To verify our assignments from the *ab initio* DFT calculations, we also performed RIXS measurements on other single-valence manganites for comparison. Figure 5b-d displays the RIXS intensity map (upper panels) and the associated XAS spectra (bottom panels) for MnO (purely $Mn^{2+}$), $LaMnO_3$ (purely $Mn^{3+}$), and $CaMnO_3$ (purely $Mn^{4+}$), respectively. These stoichiometric single-valence manganites and $LCMO_{2.5}$ share similar insulating and antiferromagnetic properties (Figures S8 and S9, Supporting Information). Notably, the XAS spectrum of $LCMO_{2.5}$ closely resembles that of MnO in Figure 5b, verifying a significant amount of $Mn^{2+}$ in $LCMO_{2.5}$. Importantly, the RIXS intensity map of $LCMO_{2.5}$ exhibits Raman-like features between 2 and 6 eV, similar to those observed for MnO, which reflects the localized electronic bands (Figure 3b) and supports its strong insulating behavior (Figure 2b,c). However, several differences can be highlighted: (i) a broad peak between 1 and 2 eV appears in $LCMO_{2.5}$ but is absent in MnO; (ii) a fluorescence-like feature above 6 eV is observed in $LCMO_{2.5}$ but not in MnO; (iii) At an incident energy of $E_i$ = 640 eV, the spectral weight is significantly enhanced in $LCMO_{2.5}$ compared to MnO.

The differences (i) and (ii) of the RIXS spectra between $LCMO_{2.5}$ and MnO can be attributed to the presence of the $Mn^{3+}$ valence state in $LCMO_{2.5}$. Figure 5c shows the RIXS intensity map of $LaMnO_3$, where $Mn^{3+}$ ions occupy only octahedra sites. A peak between 1 and 2 eV exists in Figure 5c, similar to the broad peak in $LCMO_{2.5}$ at roughly the same energy loss. In contrast, this peak is absent in both MnO ($Mn^{2+}$) and $CaMnO_3$ ($Mn^{4+}$) with octahedral configurations (Figure 5b,d), demonstrating that it originates from *dd* excitations of the $Mn^{3+}$ in an octahedral crystal field. The difference (iii) indicates that the peak at 640 eV in the bottom panel of Figure 5a is mainly contributed to tetrahedral $Mn^{2+}$, as MnO possesses only octahedral coordination.

The spectral weight of the fluorescence-like feature above 6 eV in $LCMO_{2.5}$ is suppressed compared to that of $LaMnO_3$. This feature directly associates with the O 2*p* and Mn 3*d* hybridization, as it displays significant non-local character. The suppression of spectral weight aligns with the overall weaker $Mn^{3+}$ 3*d* - O 2*p* hybridization in $LCMO_{2.5}$. Notably, this feature resonates at the lower incident energy compare to its resonance profile in $LaMnO_3$. Overall, the RIXS map agrees well with the peak assignments in XAS based on DFT calculations (Figure 5a).

Another question is whether $LCMO_{2.5}$ contains $Mn^{4+}$. Figure 5d displays the RIXS intensity map of $CaMnO_3$, which contains only octahedrally coordinated $Mn^{4+}$. Below $E_i \approx 640.5$ eV, the RIXS intensity is extremely weak, unlike that observed in $LCMO_{2.5}$. Additionally, $CaMnO_3$ exhibits a strong and broader fluorescence-like signature above 5 eV, with most of its spectral weight resonating at incident energies above 641 eV. These differences between $LCMO_{2.5}$ and $CaMnO_3$ suggest that the amount of $Mn^{4+}$ in $LCMO_{2.5}$ is minimal, confirming that most of $Mn^{4+}$ is reduced to $Mn^{2+}$ during the reduction process, as the $d^3$ and $d^5$ electronic configurations are equally stable in the octahedral crystal field.

### D. Assignments of *dd* excitations with orbital- and spin-resolved DOS calculations in $LCMO_{2.5}$

Furthermore, we analyze the *dd* excitations in $LCMO_{2.5}$ to reveal the relevant energy scales associated with Jahn-Teller distortion, Coulomb repulsion, electron hopping, and Hund's coupling. We compare the RIXS spectra of $LCMO_{2.5}$, MnO, and $LaMnO_3$ by integrating the RIXS intensity map along the incident energy axis, as depicted in **Figure 6**a. Overall, we can divide the RIXS spectrum of $LCMO_{2.5}$ into two distinct energy loss regimes. Below 3 eV, the RIXS spectrum exhibits broader peaks that resemble those observed in $LaMnO_3$. Therefore, these features are closely related to *dd* excitations of octahedrally coordinated $Mn^{3+}$ ions. Above 3 eV, the peaks become sharper and align well with the peak energies of MnO, and are mainly due to *dd* excitations of $Mn^{2+}$ in the octahedral site. Furthermore, we also note that the features above 4 eV are broader compared to the sharper peaks between 3 - 4 eV, and the peak at ≈ 4.3 eV is absent in the RIXS spectra of MnO. These observations suggest that the peaks above 4 eV are not solely attributed to *dd* excitations of $Mn^{2+}$ in the octahedral site; instead, they result from a mixed contribution from *dd* excitations of both tetrahedrally and octahedrally coordinated $Mn^{2+}$ ions.

The assignments derived from comparing the RIXS spectra of various manganites are consistent with our DFT calculations. Figure 6b displays the orbital- and spin-resolved DOS for the Mn 3*d* orbitals. The calculated bandgap is about 1.25 eV, defined as the energy difference between the $Mn^{2+}$ $e_g$ spin-up state below the Fermi level and the $Mn^{3+}$ $e_g$ spin-up state above the Fermi level. This DOS calculation aligns well with the RIXS spectrum of $LCMO_{2.5}$, where the spectral weight is nearly zero below 1.25 eV. At slightly higher energy in the conduction band, a strong peak centered about 2 eV is present, corresponding to the $Mn^{3+}$ $t_{2g}$ spin-down state. These empty $Mn^{3+}$ $e_g$ spin-up and the $t_{2g}$ spin-down states below 2 eV allow excitations from other sites to the $Mn^{3+}$ site, which accounts for the relatively broad peaks observed below 3 eV in the $LCMO_{2.5}$ spectrum. Above 2 eV, the DOS is dominated by the spin-down states from $Mn^{2+}$ sites due to the strong on-site Coulomb interaction $U$.

Consequently, the broad interband component of the RIXS spectra above 3eV is related to spin majority to minority interband transitions at $Mn^{2+}$ sites. The spin-down channel of $Mn^{2+}$ in the tetrahedral site appears around 3.5 eV in the conduction band, which partially contributes to excitations above 4 eV. Thus, despite the complexity arising from the presence of multiple valences and crystal fields in $LCMO_{2.5}$, our DFT calculations are in good agreement with the RIXS results.

Figure 6c highlights a refined peak assignment for $LCMO_{2.5}$ based on a combination of experimental data from $LCMO_3$ and DFT calculations. The convoluted joint DOS in Figure 6c, calculated from DFT over the -3 - 4 eV range (Figure 6b), is plotted for phenomenological comparison. While a direct qualitative comparison between RIXS intensity and the joint DOS is not feasible, the peak energies are generally well captured by this method. Nevertheless, we found that the peak assignments below 3 eV require modification, as they do not solely originate from $Mn^{3+}$ on-site *dd* excitations due to the presence of occupied $Mn^{2+}$ states in the spin-up channel near the Fermi level. Given the proximity of the $Mn^{2+}$ $e_g$ and $t_{2g}$ states to the valence band maximum relative to the $Mn^{3+}$ $e_g$ state, excitations observed between 2 and 3 eV should be assigned to on-site *dd* excitations of $Mn^{3+}$ within the octahedral site. In contrast, excitations below 2 eV arise from the intra-layer (within the $O_h$ layer) and inter-layer (from $T_d$ to $O_h$ layer) electron hopping from $Mn^{2+}$ to $Mn^{3+}$. This observation indicates that in $LCMO_{2.5}$ the electron kinetic energy is lower than the splitting energy of the $Mn^{3+}$ $e_g$ bands induced by Jahn-Teller distortion. Notably, the peak between 1 and 2 eV shifts to a lower energy in $LCMO_3$ (Figures S10 and S11, Supporting Information), where it is associated with *dd* excitations within the $Mn^{3+}$ $e_g$ bands.[39,40] In this regard, the Jahn-Teller distortion at the $Mn^{3+}$ site is enhanced in $LCMO_{2.5}$ due to topotactic reduction.

From above, our experimental and theoretical results demonstrate a mixed Mn valence state of $Mn^{2+}/Mn^{3+}$ in $LCMO_{2.5}$, with DFT calculations predicting a high-spin state for both octahedral and tetrahedral Mn sites. Within the tetrahedral layer, all Mn ions exhibit a $Mn^{2+}$ state due to the structural transition, consistent with previous studies on brownmillerite half-doped manganites.[29] The ordered oxygen vacancies induced in these tetrahedral layers elongate the Mn–O–Mn bonds and weaken Mn–O hybridization. As a result, the kinetic energy gain of electrons is insufficient to overcome the potential energy cost from on-site Coulomb repulsion, leading to strong electron localization. This localized behavior is reflected in the sharp peaks observed in the *dd* excitations of both octahedral and tetrahedral $Mn^{2+}$ sites. Notably, the tetrahedral chains are oriented only along the [010] direction, inducing significant structural and transport anisotropy, as supported by DFT calculations of the energy band shown in Figure 3c. At the *Γ*-point, the ratio between effective masses along the [100] and [010] directions is estimated to be $m^*_{[100]}/m^*_{[010]} = 2.12$.

However, unlike half-doped systems, where both $Mn^{2+}/Mn^{4+}$ and $Mn^{2+}/Mn^{3+}$ configurations can coexist in the octahedral layer, the one-third doping at the A sites in $LCMO_{2.5}$ precludes the formation of an ordered $Mn^{2+}/Mn^{4+}$ configuration. Instead, the $Mn^{2+}/Mn^{3+}$ configuration predominates. Similarly, the uniform oxygen deficient phase, $La_{0.7}Sr_{0.3}MnO_{2.65}$, which exhibits insulating behavior and antiferromagnetism, also belongs to the $Mn^{2+}/Mn^{3+}$ configuration.[24] The strong Jahn-Teller distortion at the $Mn^{3+}$ sites, as evidenced by the splitting of spin-up $e_g$ orbitals (Figure 6b), will trap electrons and result in super-exchange-driven antiferromagnetic interactions between octahedral $Mn^{2+}$ and $Mn^{3+}$ ions.[48] This is distinct from the ferromagnetic coupling in $LCMO_3$ films, where itinerant electrons mediate double-exchange interactions between $Mn^{3+}$ and $Mn^{4+}$ ions.[23,24,45,46]

The approach we developed to distinguish *dd* excitations in brownmillerite films—identifying the valence ion and associated coordination crystal field from which the multiplet originates—can be extended to other functional materials with multiple site occupancies and oxidation states. When perovskite oxides are topotactically reduced, the oxygen stoichiometry can be precisely controlled in a block-by-block manner, gradually transitioning from unit-cell-level superlattices of *N* octahedral layers and one tetrahedral layer (*N*:1) to $ABO_{2.75}$, $ABO_{2.67}$, and $ABO_{2.5}$, corresponding to 3:1, 2:1, and 1:1 ratios, respectively.[1,2,49,50] Furthermore, oriented variants of oxygen deficiency can be modulated in films, allowing alternating oxygen octahedra and tetrahedra to align not only along the stacking direction but also along the in-plane direction or at specific angles relative to the substrate normal. Therefore, this approach can provide valuable insights into uncovering the electronic structure of a broad range of functional materials with complex coordinated crystal fields and/or possible multiple valence states.[6-28,50-53]

## 3. Conclusions

In summary, coherently epitaxial $LCMO_{2.5}$ thin films with a brownmillerite structure were topotactically reduced from their perovskite $LCMO_3$ counterparts, undergoing a phase transition from a ferromagnetic metallic to an antiferromagnetic insulating ground state. This reduction induces mixed $Mn^{2+}/Mn^{3+}$ oxidation states and alternating octahedral and tetrahedral crystal fields along the growth direction in $LCMO_{2.5}$. The element- and site-selective capabilities of RIXS further enable the determination of relevant energy scales associated with Jahn-Teller distortion, electron hopping, Coulomb repulsion, and Hund's coupling, thereby advancing the microscopic understanding of the complex interplay between strongly correlated degrees of freedom in brownmillerite manganites. As such, the $LCMO_3$-to-$LCMO_{2.5}$ transition involves the combined effects of orbital localization, hybridization, and reconfiguration, structural transformation, and modifications in magnetic interactions.

Our work will not only advance the fundamental understanding of the materials like manganites but also lay the groundwork for developing emerging functional materials with multiple oxidation states and complex coordination crystal fields. Such materials hold significant potential for technological applications in areas like spintronics, ionic conductivity, anion insertion applications, and advanced sensors, where precise control over electronic and magnetic properties is essential for next-generation devices.

## 4. Experimental Section

*Sample Fabrication:* Using a customized pulsed laser deposition system equipped with a KrF excimer laser at a wavelength of 248 nm, $LCMO_3$, $LaMnO_3$, $CaMnO_3$ perovskite oxide thin films were deposited on NGO(001) substrates of size 5 mm*3 mm*0.5 mm. The deposition was carried out at an energy density of approximately 2 J/cm$^2$ and a repetition rate of 2 Hz. The sample-to-target distance was fixed at 5.5 cm. During deposition, the substrate temperature was maintained at 735 °C with an oxygen pressure of 40 Pa. Before cooling down to room temperature, all samples were in situ annealed for 15 mins. All films have a thickness of 24 nm. After obtaining perovskite films, some $LCMO_3$ films were individually wrapped in aluminum foil and sealed together with 1 g of calcium hydride ($CaH_2$) powder in a glass tube. This aluminum-wrapped package was then heated to 500 °C and maintained at this temperature for 15 hours to transform the crystal structure from perovskite $LCMO_3$ to brownmillerite $LCMO_{2.5}$ (Table S2, Supporting Information). Meanwhile, the other $LCMO_3$, $LaMnO_3$, and $CaMnO_3$ films were annealed in a flowing $O_2$ atmosphere at 600 °C for 3 hours to eliminate any oxygen vacancies formed during fabrication. Notably, the MnO sample used for comparison here is a single crystal.

*Structural, magnetic, and electric characterizations:* Structural characterizations of these samples were performed using XRD with Cu $K\alpha1$ radiation (PANalytical Empyrean), utilizing $2\theta$-$\omega$ linear scans and rocking curves. Their epitaxial strain states were further characterized by reciprocal space mappings. Temperature-dependence magnetization (*M*-*T*) and magnetic hysteresis (*M*-*H*) measurements were performed using a Quantum Design magnetic property measurement system (MPMS3). The magnetic field was applied parallel to the [010] direction. Resistivity versus magnetic field ($\rho$-*H*) curves were determined by a standard four-point probe method with a Quantum Design physical property measurement system (PPMS). Prior to the measurements, stripe-shaped Pt electrodes were patterned onto the sample surface using photolithography.

*STEM measurement:* The specimen of $LCMO_{2.5}$ was fabricated using a Thermal Fisher Helios G5x Dual Beam system with a focused ion beam. Before ion beam milling, a W/C layer was deposited on the top surface to protect the sample. A $Ga^+$ ion beam at 30 kV was then used to acquire the lamella. The lamella was further thinned to approximately 40 nm, followed by the use of $Ga^+$ ion beams at 5 and 2 kV acceleration voltages to remove the amorphous layers. Atomic resolution HADDF-STEM images were obtained on a Thermal Fisher Themis Z microscope, equipped with a probe-forming aberration corrector and operated at 300 kV, with a semi-convergence angle of approximately 25 mrad.

*XAS and RIXS measurements:* XAS and RIXS experiments at the Mn-*L* edge were performed at the 2ID-SIX beamline at NSLS-II, Brookhaven National Laboratory (USA). The XAS spectra were collected at 40 K in partial fluorescence yield (PFY) by integrating both of the elastic and inelastic parts of the RIXS signal as a function of incident x-ray energy. The incident x-ray is $\sigma$-polarized. The $\sigma$-polarization refers to the electric field vector of the light being perpendicular to the scattering plane. The grazing angle $\theta'$ is defined as the angle between the incident x-ray and the sample surface, and $2\theta$ is defined as the angle between the incident and scattering x-rays. The energy resolution of RIXS at the Mn *L* edge is about 21 meV, as determined by elastic measurements of a multilayer reference sample. In addition, XAS spectra of Mn-$L_{2,3}$ and O-*K* edges in Figure 4 were performed in the total-electron-yield (TEY) mode at the BL08U1A beamline of the shanghai synchrotron radiation facility (SSRF) and the BL12B beamline of the Hefei National Synchrotron Radiation Facility (NSRF).

*DFT calculations:* DFT calculations are performed using Vienna Ab-initio Simulation Package (VASP)[54,55] to provide insights on the charge-ordered insulating phase in the $LCMO_{2.5}$. The revised Perdew-Burke-Ernzerhof (PBE) for solids (PBE-sol) are used to treat the exchange-correlation functional in our calculation.[56] The projector augmented wave (PAW) method is used,[57] and the following electronic configurations for elements are considered: $6s^2 5s^2 5p^6 5d^1$ for La, $2p^6 3s^2 3p^6 4s^2$ for Ca, $3p^6 4s^2 3d^7$ for Mn, and $2s^2 2p^4$ for O. A 7×3×3 $\Gamma$-centered Monkhorst-Pack *k*-point mesh is used for calculating the orthorhombic unit cell, and scale inversely for the lattice constants in the supercell calculation. Integrations are performed using Gaussian smearing with a width of 20 meV.

The electron correlation effects among the localized Mn-3*d* states are accounted for using the PBE+*U* proposed by Liechtenstein.[58] The correlation effects (*U*) of the transition metal elements Mn 3*d* and La 4*f* states are expected to affect the relaxed structure of $LaMnO_3$. The value of *U* has been reported to affect the relaxed structure parameters in manganites.[59] Therefore, we have tested the correlation *U* for both the La-4*f* and Mn-3*d* states within the PBE-sol plus Hubbard *U* framework, and the optimal parameters, i.e., $U_{(Mn)}$ = 4 eV and $U_{(La)}$ = 6 eV, have been used throughout the article.

**Supporting Information**

Supporting Information is available from the Wiley Online Library or from the author.

**Acknowledgments**

We thank Mark Dean for the fruitful discussions. This work is supported by the National Natural Science Foundation of China (Grant Nos. U2032218, 11974326, 12074365, 12374094, 12304153, 12474232, and 12274194), the National Key Research and Development Program of China (Grant Nos. 2023YFA1406404, 2020YFA0309100, and 2020YFA0308900), the Anhui Provincial Natural Science Foundation (Grant No. 2308085MA15), the open fund of Information Materials and Intelligent Sensing Laboratory of Anhui Province (Grant No. IMIS202107), the Natural Science Foundation of Guangdong Province (2021A1515110389) and the Science Technology, Innovation Commission of Shenzhen Municipality (JCYJ20210324104812034), and the Guangdong Innovative and Entrepreneurial Research Team Program (Grant No. 2021ZT09C296). The work at Brookhaven National Laboratory was supported by the DOE Office of Science under Contract No. DE-SC0012704, the Laboratory Directed Research and Development project of Brookhaven National Laboratory Nos. 21-037 and 25-022, and the U.S. Department of Energy (DOE) Office of Science, Early Career Research Program. The research used Beamline 2-ID of the National Synchrotron Light Source II, a U.S. Department of Energy (DOE) Office of Science User Facility operated for the DOE Office of Science by Brookhaven National Laboratory under Contract No. DE-395SC0012704. This work was partially carried out at the USTC Center for Micro and Nanoscale Research and Fabrication, and at the Instruments Center for Physical Science, University of Science and Technology of China.

**CONTRIBUTIONS**

F.J., S.F., P.J., and W.W. conceived the study. F.J. grew the manganite thin films and conducted the reduction process to obtain the brownmillerite LCMO2.5 with the help of Q.L., J.Z., and Z.H. F.J. and Q.L. performed the structural characterizations and XAS measurements in TEY mode with the help of Z.Z., J.L., and L.W. F.J. carried out the magnetic and electrical transport measurements. S.F. and P.J. performed the XAS measurements in PFY mode and the RIXS measurements. M.Q.G. carried out the DFT calculations. M.G.

fabricated the lamella and performed the STEM measurements. F.J., S.F., M.Q.G., P.J., and W.W. analyzed the data and wrote the manuscript. All authors discussed the results and contributed to the manuscript.

**Conflict of Interest**

The authors declare no conflict of interest.

**Data Availability Statement**

The data that support the findings of this study are available from the corresponding author upon reasonable request.

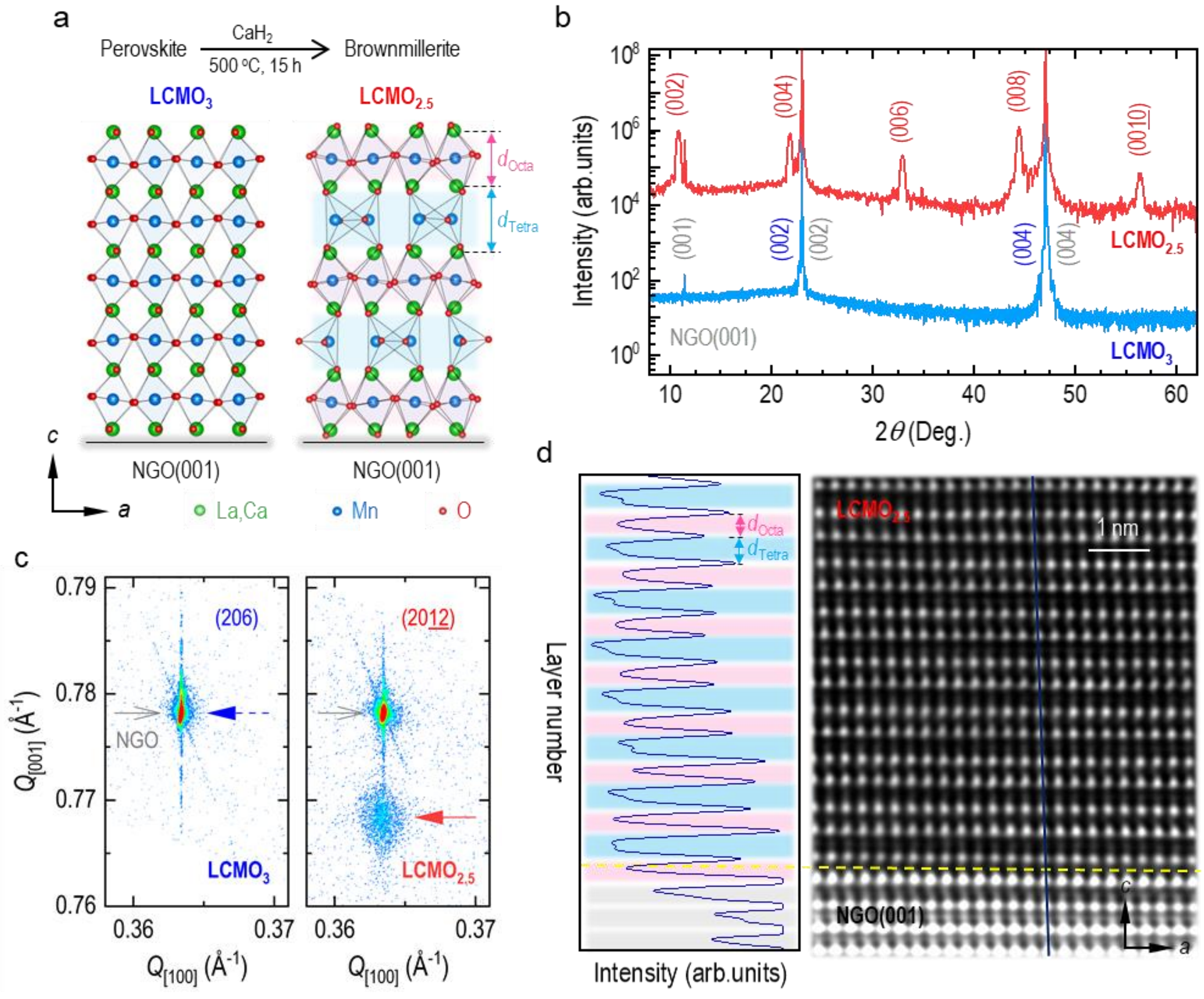

**Figure 1.** Structural characterizations of $LCMO_3$ and $LCMO_{2.5}$ films. a) Schematic diagrams of perovskite $LCMO_3$ and brownmillerite $LCMO_{2.5}$ thin films grown on NGO(001) substrates. b) XRD $2\theta$-$\omega$ linear scans and c) Reciprocal space mappings of $LCMO_3$ and $LCMO_{2.5}$ films. Due to the alternating octahedral and tetrahedral layers along the stacking direction, the *c* lattice constant of $LCMO_{2.5}$ is nearly twice that of $LCMO_3$, yielding an additional set of reflections and a doubling of the index along the [001] direction. d) The left panel: Intensity profile along the dark blue line in the right panel, showing alternating large $d_{Tetra}$ and small $d_{Octa}$ lattice spacings of the $LCMO_{2.5}$ film. The light red and blue lateral layers represent the octahedral and tetrahedral layers, respectively. The right panel: HAADF STEM image of the $LCMO_{2.5}$ film, which exhibits a sharp heterointerface (the dashed yellow line) between $LCMO_{2.5}$ and NGO.

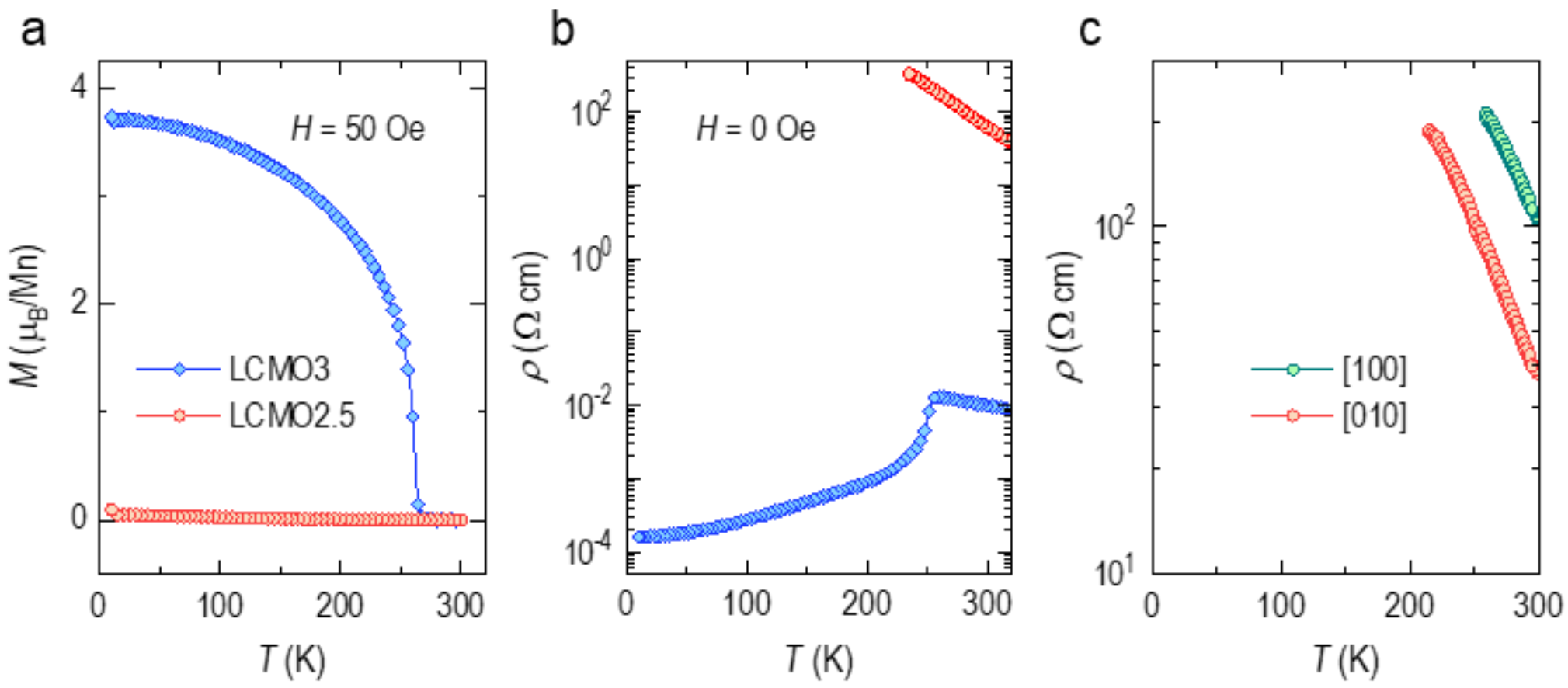

**Figure 2.** Physical properties of $LCMO_3$ and $LCMO_{2.5}$ films. a) Magnetization and b) Resistivity as a function of temperature measured from $LCMO_3$ and $LCMO_{2.5}$ films, with the resistivity exceeding the measurement range for $LCMO_{2.5}$ below approximately 250 K. The magnetization was measured at a magnetic field of 50 Oe along the [010] direction. c) Temperature-dependent anisotropic resistivity of $LCMO_{2.5}$.

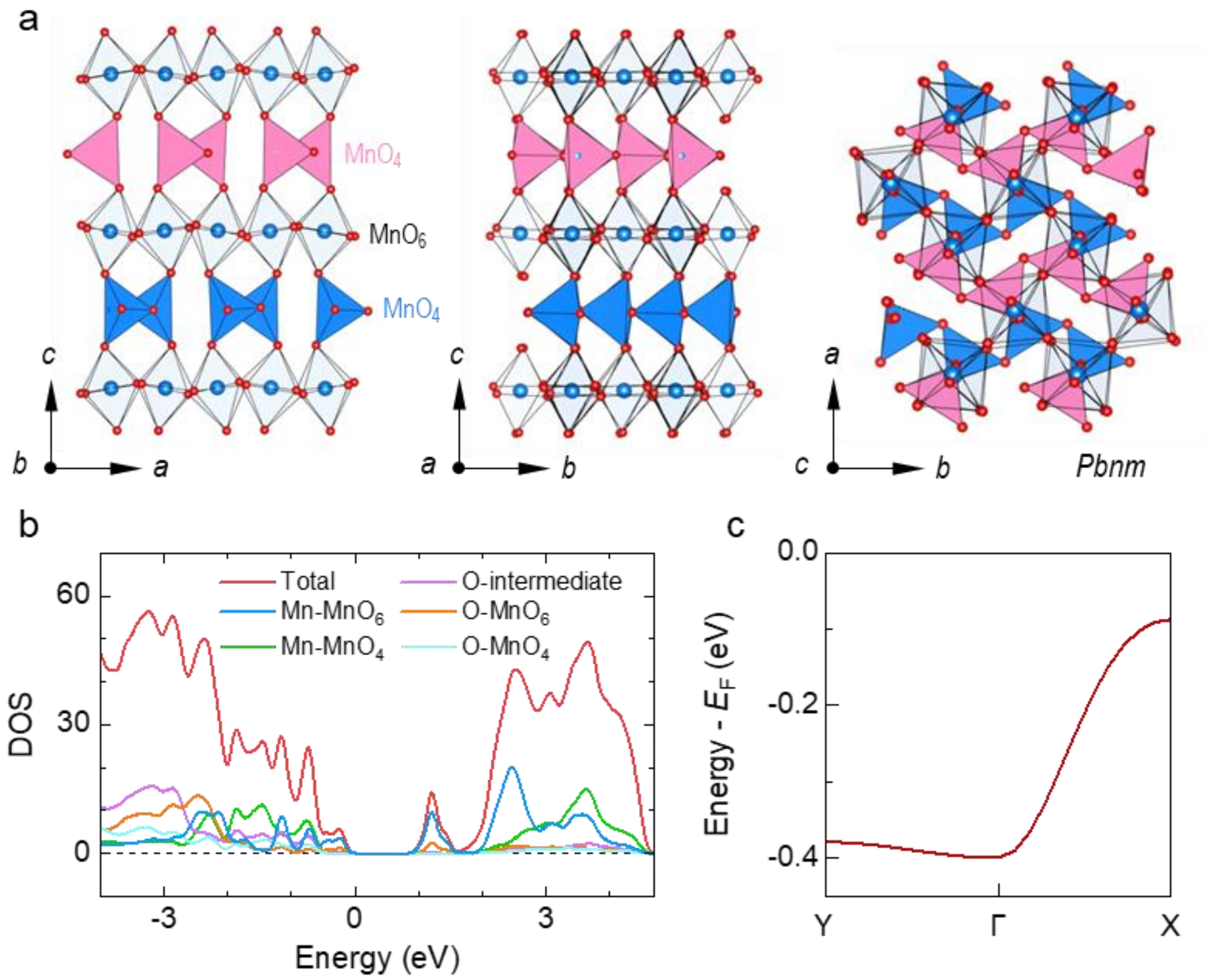

**Figure 3.** Schematics and DFT calculations of $LCMO_{2.5}$. a) Schematics of the *Pbnm* space group viewed along three different crystalline directions, illustrating the alternating ordering of tetrahedral chains between adjacent tetrahedral layers. For clarity, the A-site cations are omitted from these schematics. b) DOS as a function of energy for $LCMO_{2.5}$, showing contributions from tetrahedral and octahedral manganese ions, and intermediate, tetrahedral, and octahedral oxygen ions. c) Energy band of $LCMO_{2.5}$.

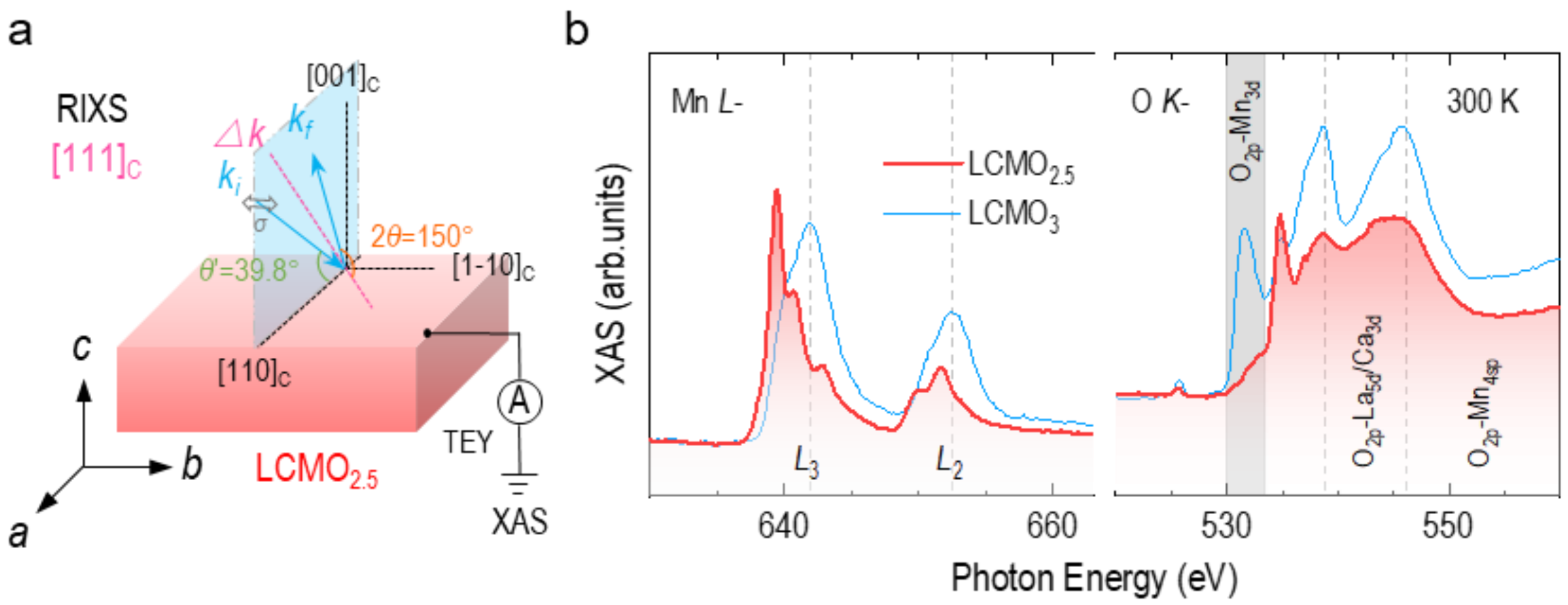

**Figure 4.** XAS of $LCMO_3$ and $LCMO_{2.5}$ films. a) Measurement geometry for XAS and RIXS. $k_i$ and $k_f$ refer to the incident and scattered x-rays, respectively. The scattering angle $2\theta$ is defined as the angle between $k_i$ and $k_f$. The grazing angle $\theta'$ is the angle between $k_i$ and the sample [110] direction. $\Delta k$ is the momentum transfer direction, parallel to the [$h$,$h$,$h$] crystallographic direction with $h$ = 0.22 r.l.u. $\sigma$ refers to the polarization of the incident x-ray. The crystal axes are defined using the pseudo-cubic notation. b) XAS spectra for $LCMO_{2.5}$ and $LCMO_3$ films at the Mn-$L$ and O-$K$ edges, collected with the total electron yield. The red and blue solid lines represent the spectra of $LCMO_{2.5}$ and $LCMO_3$, respectively.

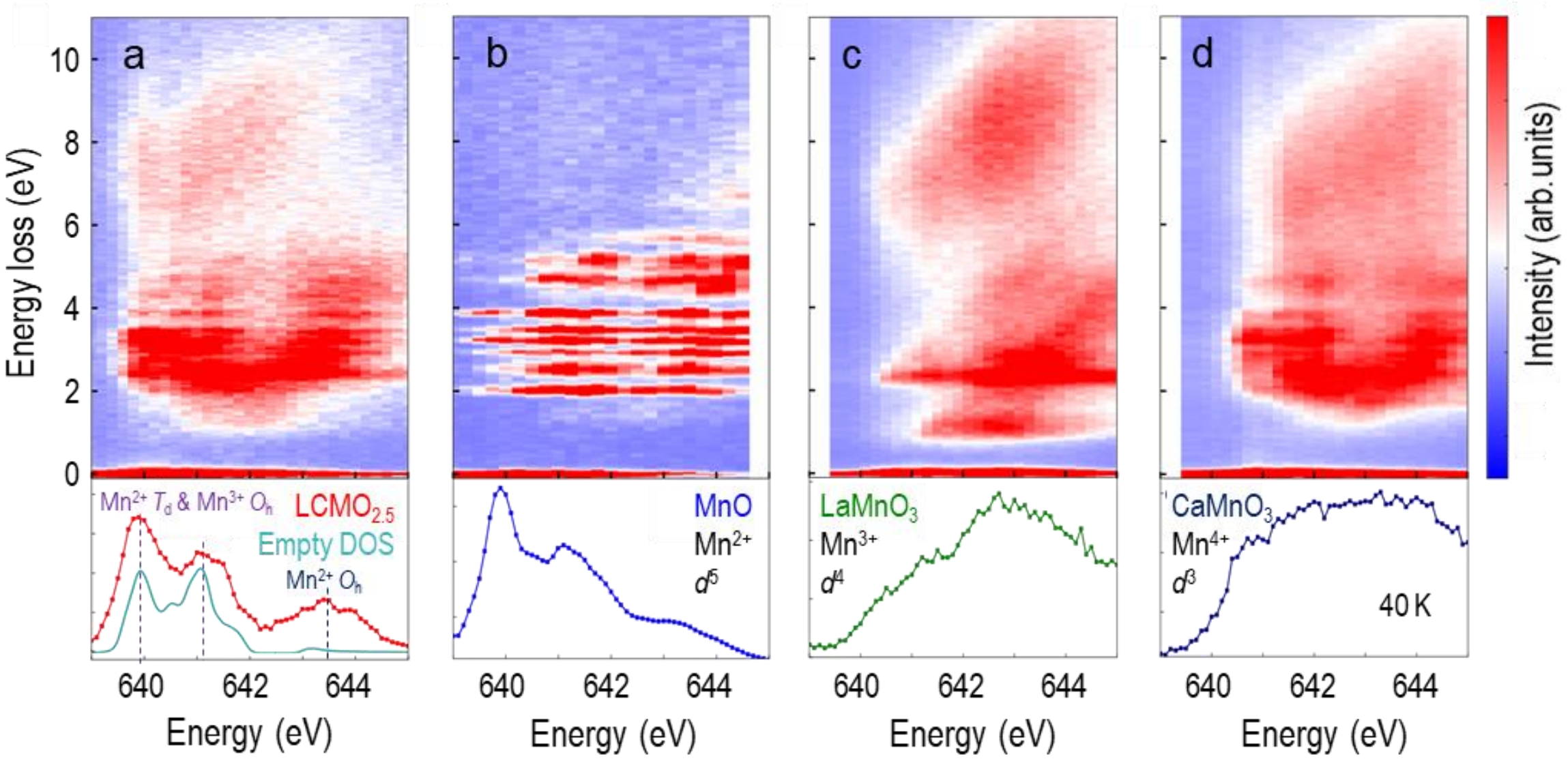

**Figure 5.** RIXS and XAS of multiple-valence brownmillerite and single-valence perovskite manganite films. Upper panels: RIXS intensity color map as a function of the incident x-ray energy and energy loss of a) $LCMO_{2.5}$, b) MnO, c) $LaMnO_3$, and d) $CaMnO_3$. All the maps are normalized to excitations from 0.4 to 12 eV for better comparison. Bottom panels: XAS of these samples at the Mn $L_3$ edge between 639 - 645 eV, using the partial fluorescence yield. The electronic configurations of MnO, $LaMnO_3$, and $CaMnO_3$ are $d^5$, $d^4$, and $d^3$, respectively. All measurements were taken at 40 K with the $\sigma$-polarization.

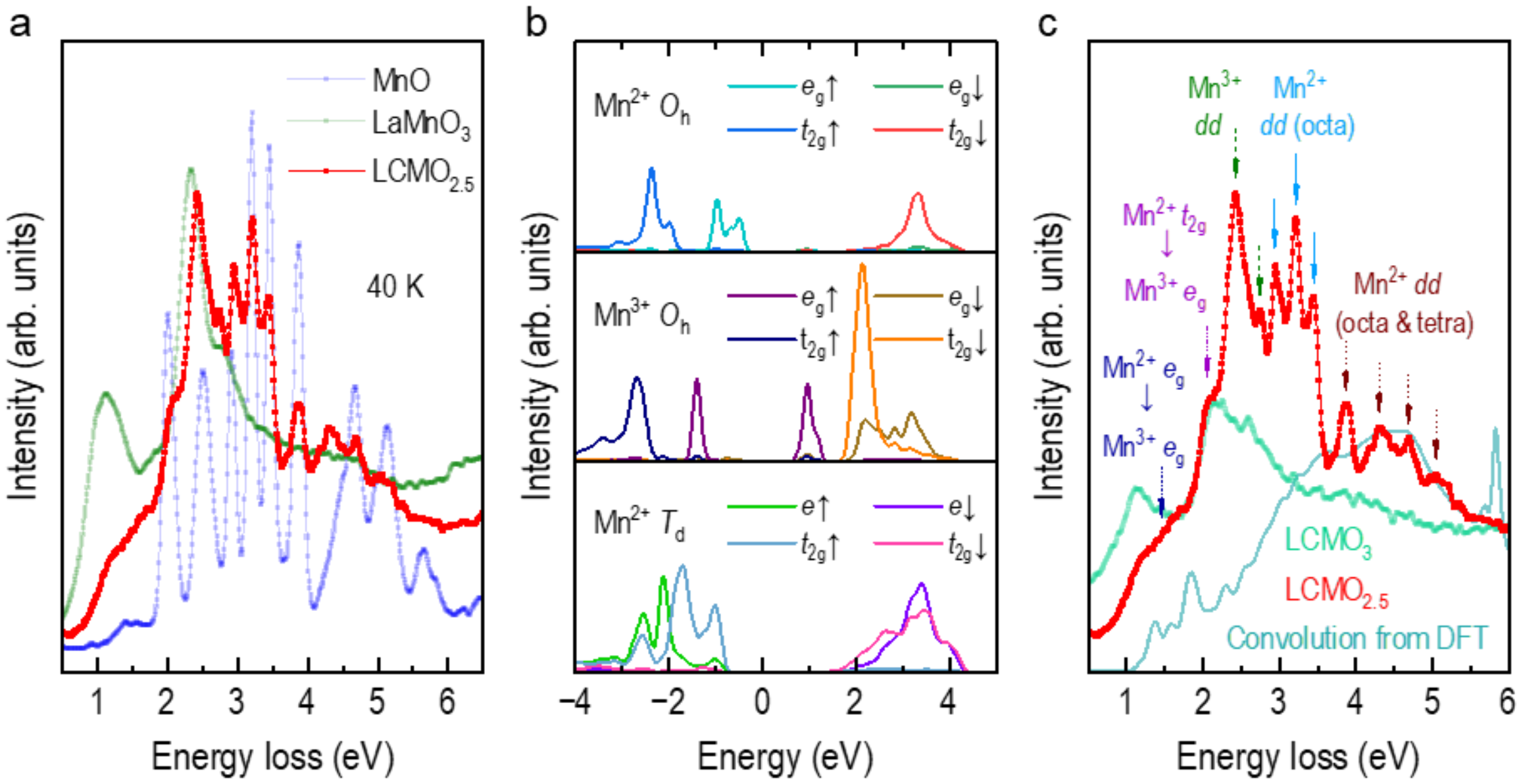

**Figure** 6. Orbital-specific crystal field excitations in $LCMO_{2.5}$ films. a) Comparison of RIXS spectra between $LCMO_{2.5}$, MnO, and $LaMnO_3$. All RIXS spectra are obtained by integrating the RIXS energy map along the incident energy axis, and are normalized to the energy loss range of 0.4 - 12 eV. The intensity of MnO is further re-scaled by a factor of two for clarity. b) Orbital- and spin-resolved DOS calculation of Mn 3*d* orbitals in $LCMO_{2.5}$, projected onto octahedral ($O_h$) $Mn^{2+}$, octahedral $Mn^{3+}$, and tetrahedral ($T_d$) $Mn^{2+}$ sites (top to bottom). c) Comparison of the RIXS spectra of $LCMO_{2.5}$ and $LCMO_3$. The convolution of the DOS of valence and conduction bands (from -3 to 4 eV) is also added for comparison. All RIXS measurements were taken at 40 K.

# Supporting Information:

# Topotactic Reduction-Driven Crystal Field Excitations in Brownmillerite Manganite Thin Films

Feng Jin[1,#] Shiyu Fan[2,#,*] Mingqiang Gu[3,4,#] Qiming Lv,[1] Min Ge,[1] Zixun Zhang,[1] Jinfeng Zhang,[1] Jingdi Lu,[1] Taehun Kim,[2] Vivek Bhartiya,[2] Zhen Huang,[5,6] Lingfei Wang,[1] Valentina Bisogni,[2] Jonathan Pelliciari,[2,†] and Wenbin Wu[1,5,7,‡]

[1]*Hefei National Research Center for Physical Sciences at the Microscale, University of Science and Technology of China, Hefei 230026, China*
[2]*National Synchrotron Light Source II, Brookhaven National Laboratory, Upton, NY 11973, USA*
[3]*Department of Physics, Southern University of Science and Technology, Shenzhen 518055, China*
[4]*School of Flexible Electronics and State Key Laboratory of Optoelectronic Materials and Technologies, Sun Yat-sen University, Shenzhen 518000, China*
[5]*Institutes of Physical Science and Information Technology, Anhui University, Hefei 230601, China*
[6]*Information Materials and Intelligent Sensing Laboratory of Anhui Province, Anhui University, Hefei 230601, China*
[7]*High Magnetic Field Laboratory, Chinese Academy of Sciences, Hefei 230031, China*

[#]These authors contributed equally to this work.
[*]sfan1@bnl.gov
[†]pelliciari@bnl.gov
[‡]wuwb@ustc.edu.cn

## Table of Contents

## Ⅰ. Strain States and Structural Characterizations of Manganites Thin Films.

[1] $LCMO_{2.5}$ and $LCMO_3$.

The $LCMO_{2.5}$, $LCMO_3$, and NGO all share an orthorhombic structure with the same *Pbnm* space group, and their lattice parameters are provided in Table S1. Compared to $LCMO_3$, $LCMO_{2.5}$ features unique tetrahedral chains along the [010] direction, interleaved by oxygen vacancies within the tetrahedral layers. Consequently, while $b_{LCMO2.5}$ is nearly identical to $b_{LCMO3}$, $a_{LCMO2.5}$ is significantly larger than $a_{LCMO3}$. Before and after the reduction process, the film always remains coherent with the NGO, resulting in lattice constants of the film "copying" those of the NGO. Using their bulk lattice constants, the strain states of $LCMO_{2.5}$/NGO(001) and $LCMO_3$/NGO(001) thin films can be determined. For $LCMO_3$/NGO(001), an anisotropic strain of +0.85% tensile strain along the [010] direction and -0.70% compressive strain along the [100] direction is induced (Figure S1a). In contrast, $LCMO_{2.5}$/NGO(001) shows a +0.24% tensile strain along the [010] direction and a −3.95% compressive strain along the [100] direction (Figure S1b).

The high quality of $LCMO_{2.5}$ and $LCMO_3$ films is demonstrated by XRD linear scans and rocking curves (Figure S2). The linear scans in Supplementary Fig.2a are magnified from Fig. 1b. Laue fringes are observed around the (008) reflection of $LCMO_{2.5}$ and the (004) reflection of $LCMO_3$, indicating atomically flat interfaces/surfaces both before and after the reduction process. Additionally, the FWHMs of these two reflections are 0.0567° and 0.0537°, respectively, further confirming the high quality of the films.

[2] $LaMnO_3$ and $CaMnO_3$.

We also performed XRD linear scans and rocking curves of $LaMnO_3$ and $CaMnO_3$ films (Figure S9a). Laue fringes are observed around the (004) reflection of the $LaMnO_3$ film, indicating atomically flat interfaces/surfaces. Additionally, the FWHMs of the (004) reflection of $LaMnO_3$ and $CaMnO_3$ films are 0.047° and 0.087°, respectively.

## Ⅱ. DFT Calculations of $LCMO_{2.5}$ and Single-Valence Manganites.

[1] Crystalline Structures and DOS of three different space groups of $LCMO_{2.5}$.

We performed the DFT simulation to reveal the arrangement for the tetrahedral chains in brownmillerite-type frameworks among the *Pbnm*, *Im2c*, and *Pmca* space groups.[47] Full ordering is achieved when described in the *Pbnm* and *Im2c* space groups, which show ordered chain arrangements in the tetrahedral layers. In the *Im2c* structure, all $MnO_4$ chains share the same orientation, whereas in the *Pbnm* structure, the orientation of $MnO_4$ chains alternates between adjacent tetrahedral layers (the middle panels of Figures 3a and S3b). By contrast, the *Pmca* space group displays an interlaced superposition of both arrangements (Figure S3a). Their crystal structures along three mutual orthogonal crystalline orientations are schematically displayed in Figures 3a and S3. For clarity, the A-site atoms are omitted. To determine the arrangement of the tetrahedral chains in our brownmillerite $LCMO_{2.5}$ films, we performed DFT calculations using a G-type antiferromagnetic structure. These three structures are energetically almost degenerate. Compared to the *Pbnm* space group, *Pmca* and *Im2c* structures are only 4 and 6 meV per formula unit (f.u.) higher in the total energy, respectively, as shown in Figure S5. Furthermore, their electronic structures near the Fermi energy, i.e. the valence band maximum and conduction band minimum, are quite similar, all of which feature a band gap of approximately 1.25 eV, as shown in Figures 3b, S3c, and S5a. As coherently epitaxial $LCMO_{2.5}$ films exert a large compressive strain, we further calculated the total energy of these three space groups as a function of in-plane biaxial strain. Figure S5c highlights that the *Pbnm* phase is always the most stable in the low-strain states.

[2] Crystalline Structures and DOS of $LCMO_{2.5}$ with varied A-site ordering.

In DFT calculations, we also consider A-site ordering for one-third-doped brownmillerite manganites. Figure S4a schematically shows various A-site orderings. Although their DFT calculations yield similar electronic structures in Figure S4b, the band gaps vary slightly. The band gaps for the *abab*-, *abca*-, and *abcb*-type A-site orderings are 1.25, 1.0, and 1.25 eV, respectively. Among these, the *abcb*-type ordering is the most stable, suggesting that the dopant Ca cations tend to isolate rather than cluster together.

[3] Valence state and schematic diagram of $LCMO_{2.5}$.

This section now includes the density of states (DOS) projected onto the $t_{2g}$ and $e_g$ orbitals for both tetrahedral and octahedral Mn sites. Figure S6b clearly shows that both $t_{2g}$ and $e_g$ orbitals for the tetrahedral Mn sites are fully occupied, indicating a +2 valence state ($d^5$). Similarly, the orbital occupation depicted in Figure S6a allows us to identify the +2 and +3 valence states for the octahedral Mn sites.

Figure S7a illustrates an ideal octahedral framework. However, both d5 and d3 states induce octahedral breathing distortions. When the ratio is 1:1, a checkerboard-type network can form, accommodating the charge-ordering pattern, as illustrated in Figure S7b. In contrast, when the ratio shifts to 2:1, the octahedral network is disrupted, resulting in non-contacting vertices and over-closed vertices, as depicted in Figure S7c.

[4] DOS of MnO, $LaMnO_3$, and $CaMnO_3$.

DFT calculations of Mn 3*d* DOS in single-valence perovskite manganites, that is, MnO, $LaMnO_3$, and $CaMnO_3$, are shown in Figure S8. The Mn 3*d* states are split into $t_{2g}$ and $e_g$ orbitals. Solid lines represent the spin-up DOS, while dashed lines represent the spin-down DOS, with the spin-up and spin-down DOS being nearly identical. The calculated band gaps for MnO, $LaMnO_3$, and $CaMnO_3$ are approximately 2.9, 0.7, and 1.2 eV, respectively. Theoretically, all these single-valence perovskite manganites are expected to have an antiferromagnetic insulating ground state.

## Ⅲ. Magnetic and Electrical Properties of Manganites Thin Films.

[1] $LCMO_{2.5}$ and $LCMO_3$.

Different ground states in $LCMO_{2.5}$ and $LCMO_3$ films are achieved only by removing oxygen through the reduction process. Although the $LCMO_3$ film exhibits a bulk-like ferromagnetic metallic ground state, the $LCMO_{2.5}$ film has an antiferromagnetic insulating ground state. As shown in Figure 2a,b, the $M-T$ curve of the $LCMO_3$ film at zero Tesla (T) shows an upturn in resistivity around 260 K, and the $\rho-T$ curve shows a downturn, indicating a phase transition from the paramagnetic insulator to double-exchange-mediated ferromagnetic metal.[8] In contrast, the $LCMO_{2.5}$ film remains insulating across the measured temperature range, with nearly zero magnetization. For the $LCMO_{2.5}$ film, measurements taken at 0 and 9 T show no significant difference in magnetotransport behavior.

[2] $LaMnO_3$ and $CaMnO_3$.

As shown in Figure S9b,c, the $\rho{-}T$ curve of the $CaMnO_3$ thin film exhibits insulating behavior across the entire measured temperature range, and the magnetization observed in the $M{-}T$ curve originates from the substrate, without any significant magnetic signal from the film.[8] These findings confirm that the $CaMnO_3$ thin film has an antiferromagnetic insulating ground state. However, for the $LaMnO_3$ thin film, the ferromagnetism appears at ~100 K, and an insulator-metal-insulator transition occurs in the $\rho{-}T$ curve, likely due to slight off-stoichiometry commonly observed in $LaMnO_3$. This off-stoichiometry may result from cation vacancies modulating the oxidation state.[9] Nevertheless, we believe that $Mn^{3+}$ ions still dominate in $LaMnO_3$.

## Ⅳ. RIXS Intensity Color Map of $LCMO_3$.

Figure S10 displays the RIXS intensity color map as a function of incident photon energy and energy loss of $LCMO_3$ at the Mn $L$ edge. Several excitations can be identified along the energy loss axis. We observe two Raman-like features below 2.2 eV and some fluorescence-like features above 2 eV. Compared to the $LCMO_{2.5}$, the strong fluorescence in $LCMO_3$ is consistent with its half-metallic behavior as the double-exchange interaction is dominated. The localized feature at about ~1.2 eV has been observed in previous optical spectroscopy, where some of the works assign it as the intersite *dd* excitation from $Mn^{3+}$ to $Mn^{4+}$.[10,11] However, from the RIXS energy map, this feature does not disperse with incident energy indicating its localized nature. Therefore, the ~1.2 eV peak should be assigned as the $Mn^{3+}$ on-site *dd* excitation. Indeed, this feature has been observed only in $LaMnO_3$ but not in $CaMnO_3$ (The upper panels of Figure 5c,d), demonstrating that the peak originates from the splitting of the $Mn^{3+}$ $e_g$ orbitals induced from the Jahn-Teller distortion as the $Mn^{4+}$ is Jahn-Teller inactive.

## Ⅴ. Fine Scans of *dd* Excitations in Manganites at Selected Incident Energies.

Figure S11 compares the line cuts of the RIXS spectrum of different manganites from the RIXS energy map (The upper panels of Figure 5 and Figure S10) at 639.8, 641, and 643.2 eV. These three different energies are associated with the three peak energies of the $LCMO_{2.5}$ XAS spectrum. The broad peak centered about 2.2 eV is clearly observed at $E_i$ = 639.8 and 641 eV. This feature is consistent with the strongest peak observed in the RIXS spectra of $LaMnO_3$, indicating that it originates from the $Mn^{3+}$ at the octahedral site. At $E_i$ = 643.2 eV, the peaks observed in the RIXS spectrum of $LCMO_{2.5}$ resemble well with those of MnO, proving that most of the peaks are related to the Mn *dd* excitations of the $Mn^{2+}$ octahedral site. Overall, our RIXS results are fully consistent with the peak assignments in the XAS of $LCMO_{2.5}$ (The bottom panels of Figure 5a in the main text).

## Ⅵ. Supplementary References.

## Ⅶ. Table, Supporting Information.

| bulk | Space group | *a* (Å) | *b* (Å) | *c* (Å) | *V* (Å$^3$) | reference |
|---|---|---|---|---|---|---|
| $LCMO_{2.5}$ | *Pbnm* | 5.6477 (-3.95%) | 5.4901 (+0.24%) | 16.140 | 500.44 | [1] |
| $LCMO_3$ | *Pbnm* | 5.4717 (-0.70%) | 5.4569 (+0.85%) | 7.7112 | 230.25 | [2] |
| $LaMnO_3$ | *Pbnm* | 5.494 (-1.12%) | 5.543 (-0.72%) | 7.805 | 237.69 | [3] |
| $CaMnO_3$ | *Pbnm* | 5.2758 (+2.90%) | 5.2781 (-4.09%) | 7.4542 | 207.57 | [4] |
| NGO | *Pbnm* | 5.4332 | 5.5034 | 7.7155 | 230.70 | [5] |
| MnO | *Fm3m* | 4.4448 | 4.4448 | 4.4448 | 87.81 | [6] |

**Table S1.** Lattice constants of manganites. Lattice constants for bulk $LCMO_{2.5}$,[1] $LCMO_3$,[2] $LaMnO_3$,[3] $CaMnO_3$,[4] NGO,[5] and MnO.[6] The calculated strain states of their corresponding films are shown in parentheses.

|  |  | Materials | Reduction temperature | References |
|---|---|---|---|---|
| 1 | Films | $Nd_{0.8}Sr_{0.2}NiO_2$<br>$LaNiO_2$ | 260-280 $^oC$ | [12] *Nature* **2019**, 572, 624–627<br>[13] *Appl. Phys. Lett.* **2009**, 94, 082102 |
| 2 |  | $SrFeO_2$<br>$SrFeO_{2.5}$ | 280 $^oC$ | [14] *Nature* **2007**, 450, 1062–1065<br>[15] *Appl. Phys. Lett.* **2008**, 92, 161911 |
| 3 |  | $Ca_3Co_3O_8$<br>$CaCoO_2$<br>$SrCoO_{2.5}$ | 200-300 $^oC$ | [16] *Nat. Mater.* **2024**, 23, 912–919<br>[17] *Nature* **2023**, 615, 237–243<br>[18] *Nat. Mater.* **2013**, 12, 1057–1063 |
| 4 |  | $La_{0.67}Sr_{0.33}MnO_{2.5}$<br>$Nd_{0.5}Sr_{0.5}MnO_{2.5}$ | 500 $^oC$ | [19] *Adv. Mater.* **2019**, 31, 1806183<br>[20] *Mater. Today Phys.* **2022**, 29, 100922<br>[21] *Acta Mater.* **2023**, 245, 118616 |
|  | Compounds | $La_{0.5}Ca_{0.5}MnO_{2.5}$<br>$La_{0.5}Sr_{0.5}MnO_{2.5}$<br>$Nd_{0.5}Sr_{0.5}MnO_{2.5}$ | 630 $^oC$ | [22] *Chem.-Eur. J.* **2008**, 14, 9038–9045 |

**Table S2.** Reduction temperatures for various materials and their associated references.

## Ⅷ. Figures, Supporting Information.

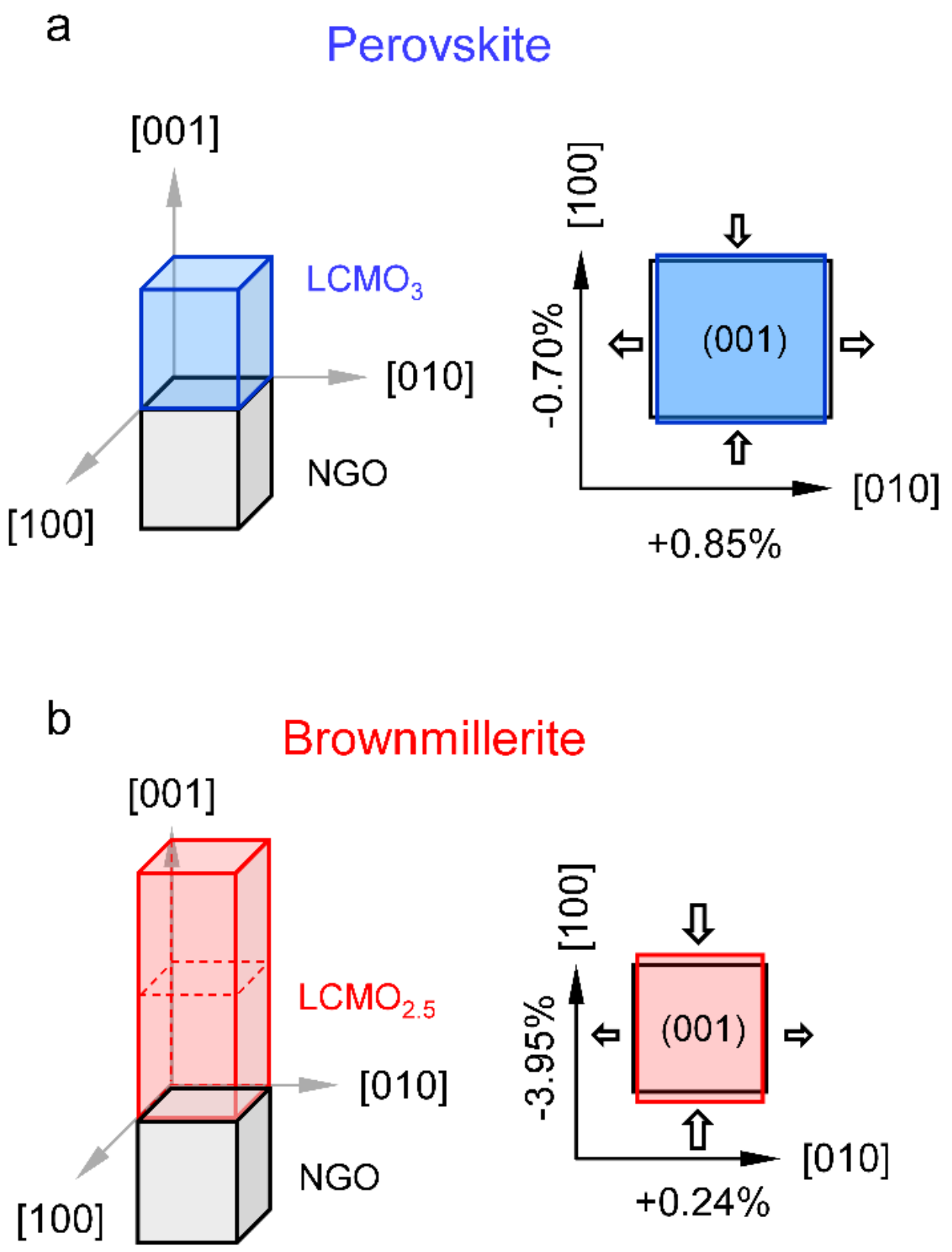

**Figure S1.** Schematics of strain states. a,b) Schematics of strain states of coherently epitaxial $LCMO_3$ and $LCMO_{2.5}$ films.

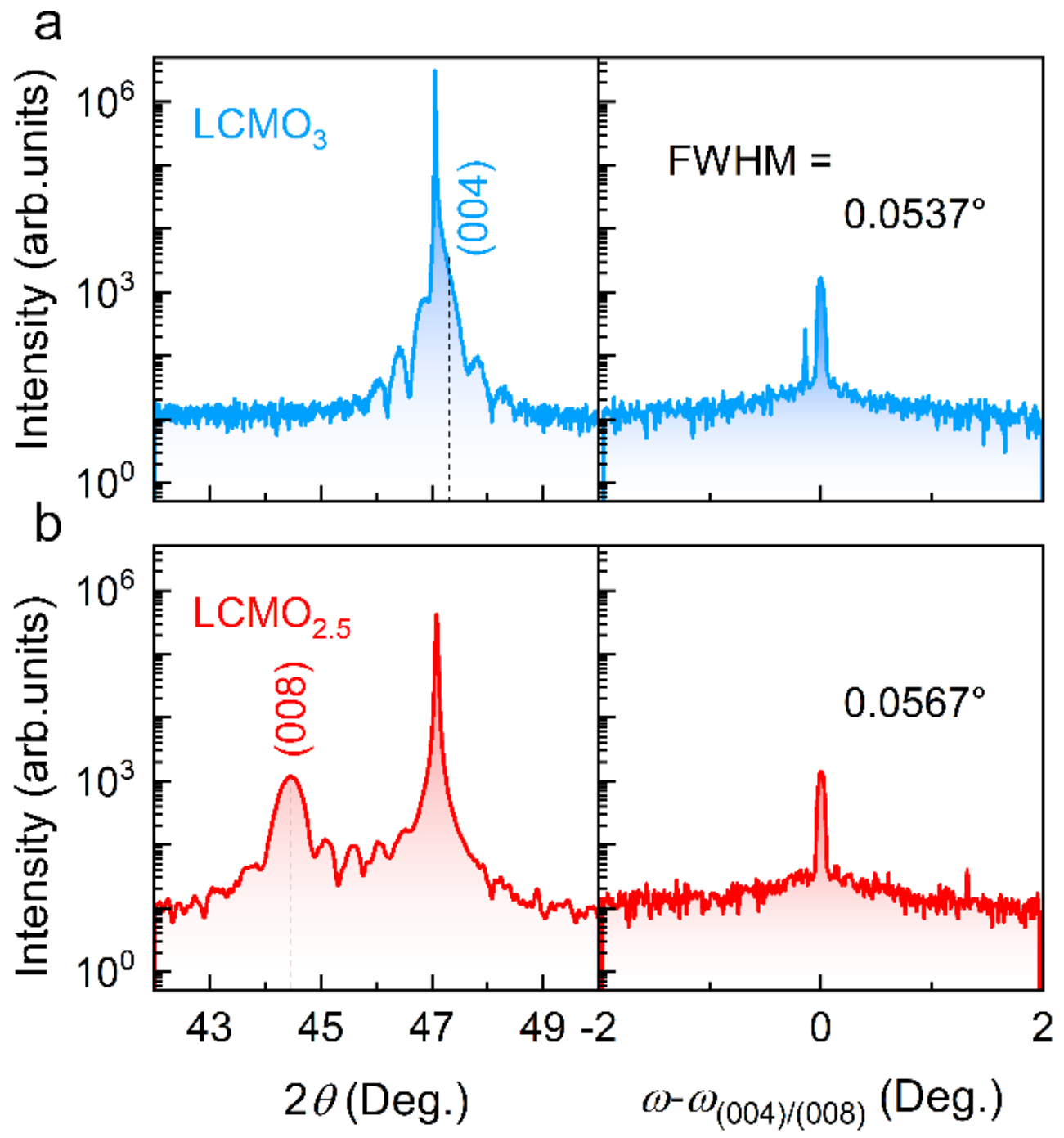

**Figure S2.** XRD linear scans and rocking curves. a) XRD linear scan (left) and rocking curve (right) around the (004) reflection of the epitaxial $LCMO_3$/NGO(001) film. b) XRD linear scan (left) and rocking curve (right) around the (008) reflection of the epitaxial $LCMO_{2.5}$/NGO(001) film. As the *c* lattice constant of $LCMO_{2.5}$ *i*s nearly double t*hat of* $LCMO_3$, the reflection index along the [001] direction is also doubled. Compared with the (004) Bragg angle of $LCMO_3$, the lattice expansion shifts the (008) Bragg angle of $LCMO_{2.5}$ to a lower angle.

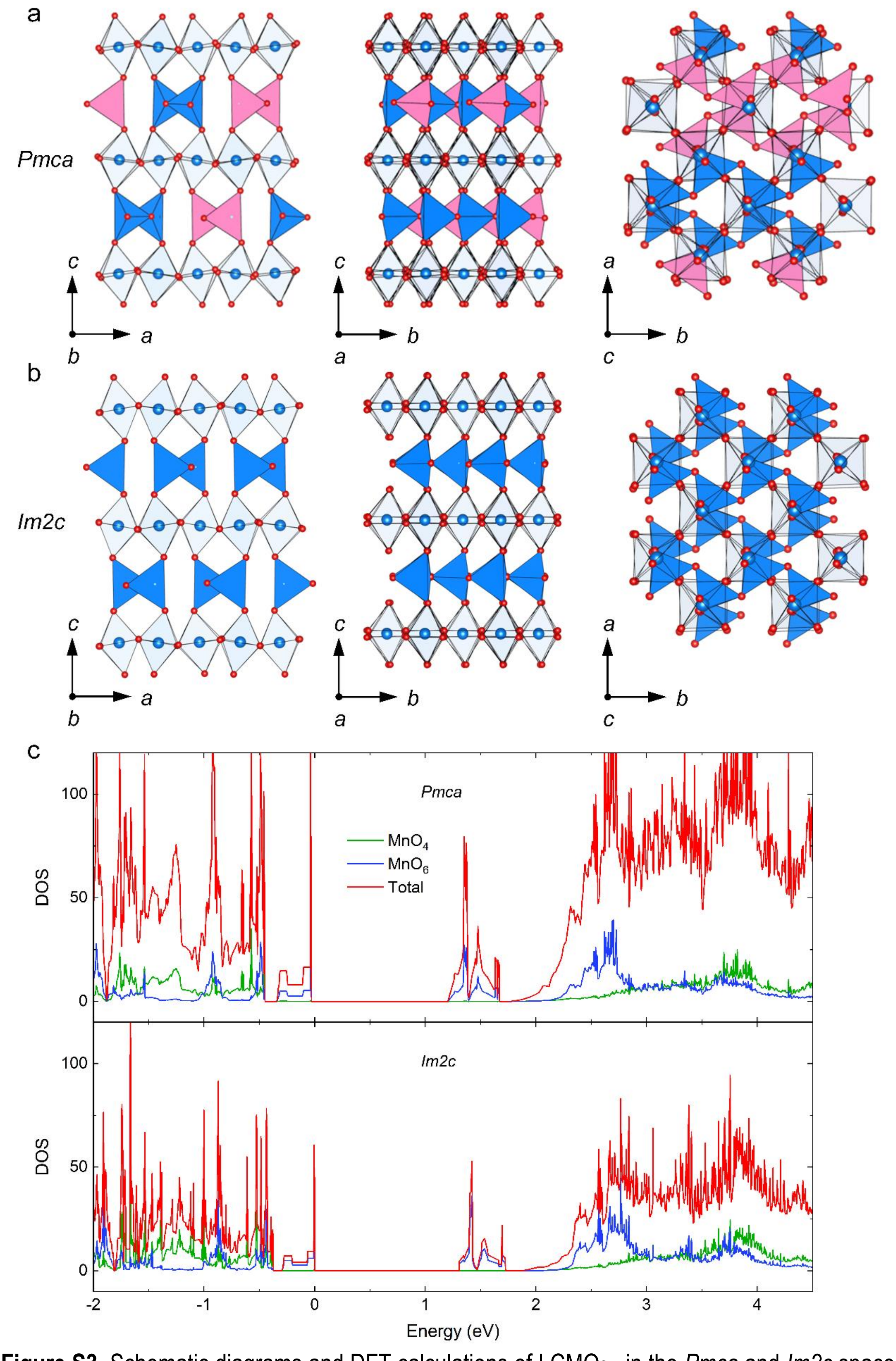

**Figure S3.** Schematic diagrams and DFT calculations of $LCMO_{2.5}$ in the *Pmca* and *Im2c* space groups. Schematic diagrams of $LCMO_{2.5}$ in the a) *Pmca*, and b) *Im2c* space groups, viewed along different crystalline orientations. For clarity, the A-site cations are omitted. c) DOS of $LCMO_{2.5}$ for the *Pmca* and *Im2c* space groups.

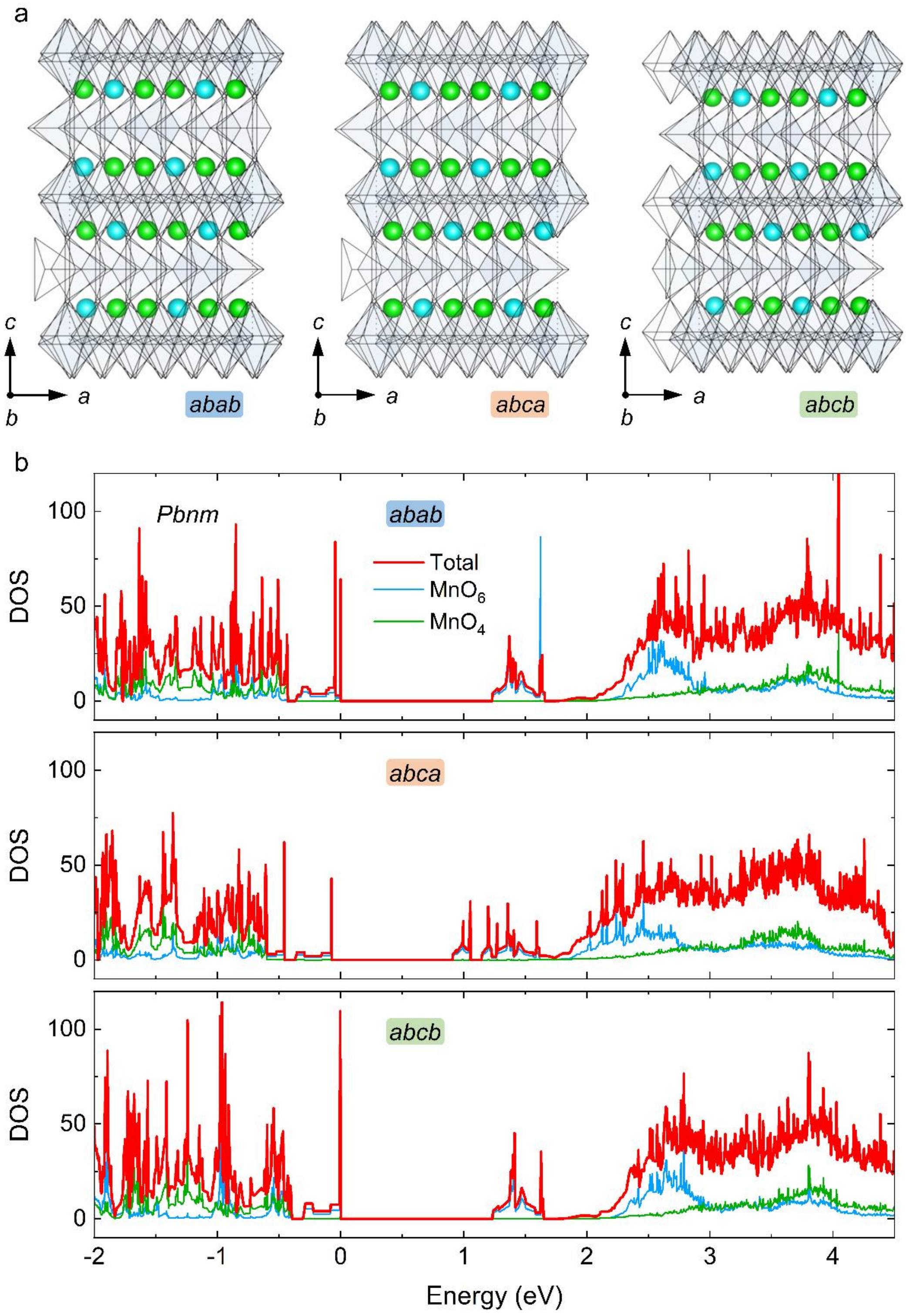

**Figure S4.** Schematic diagrams and DFT calculations of $LCMO_{2.5}$ with varied A-site ordering. a) Schematics of $LCMO_{2.5}$ with three A-site orderings in the *Pbnm* space group, where the specific configurations are shown. For clarity, the Mn and O atoms are omitted. b) DOS of $LCMO_{2.5}$ for various A-site ordering configurations.

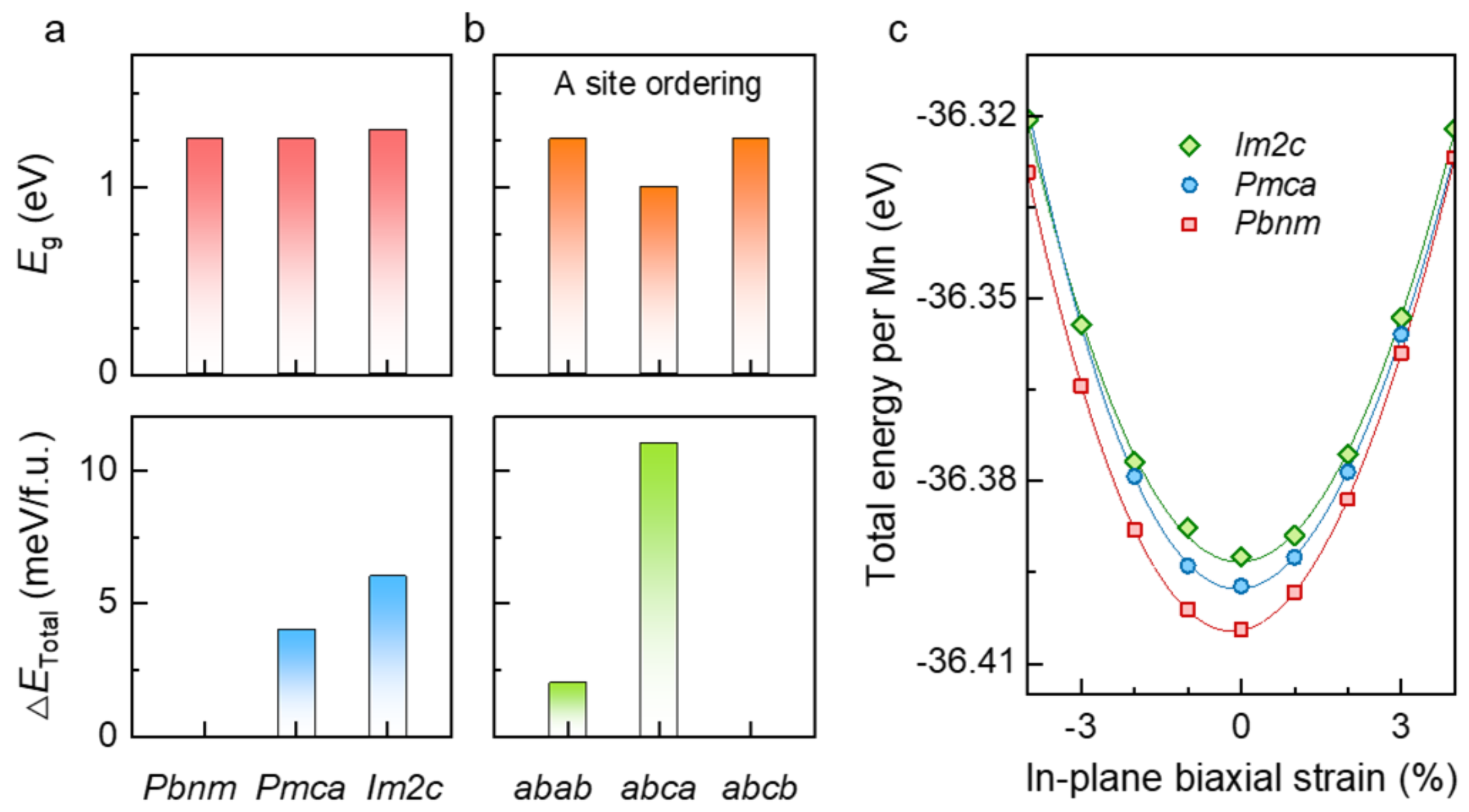

**Figure S5.** DFT calculations of $LCMO_{2.5}$ structure. a,b) Band gap ($E_g$) and relative total energy ($\Delta E_{total}$) for $LCMO_{2.5}$ in three space groups with varying tetrahedral chain arrangements or different A-site orderings. c) Total energy per Mn as a function of in-plane biaxial strain for *Im2c*, *Pmca*, and *Pbnm* space groups.

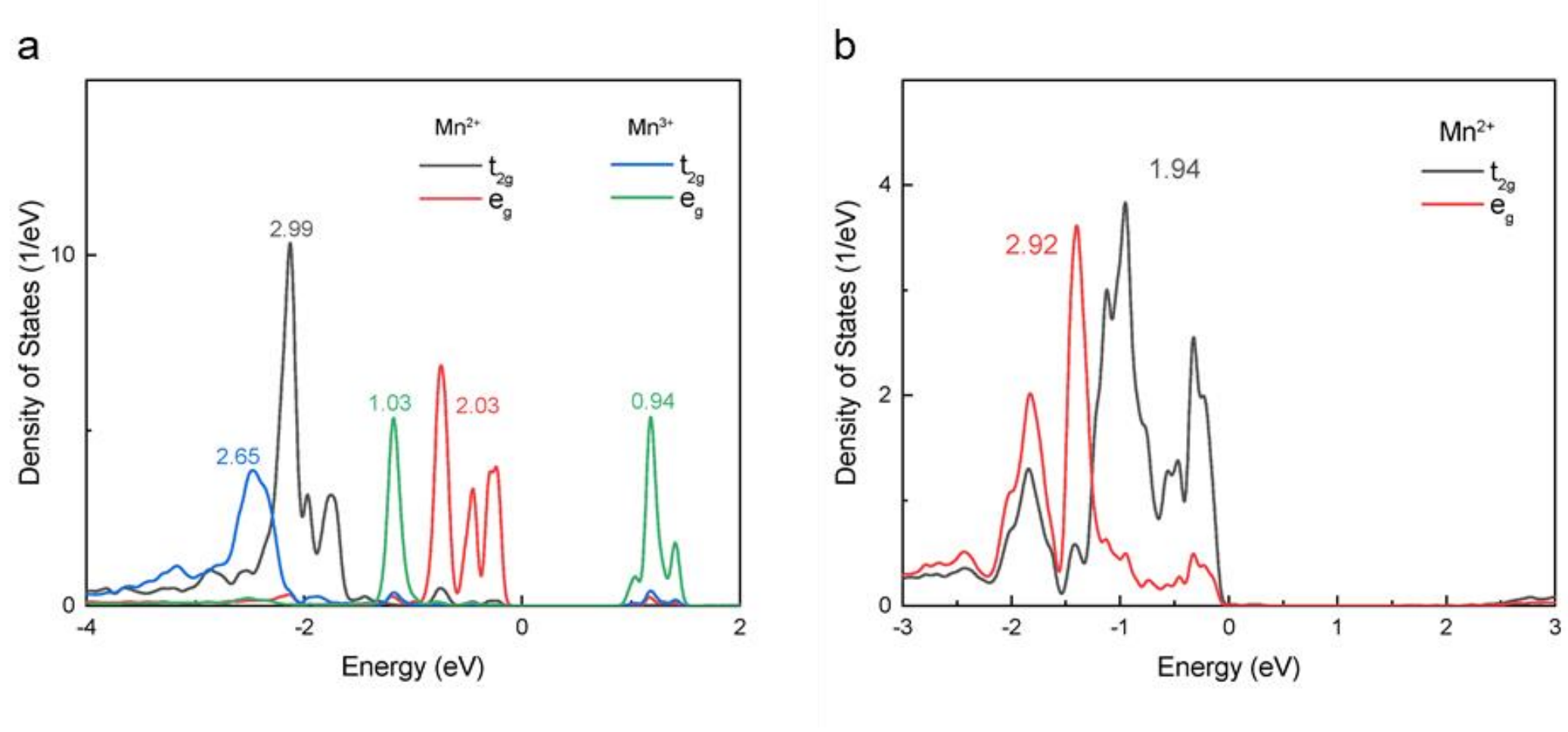

**Figure S6.** The DOS projected onto the $t_{2g}$ and $e_g$ orbitals of Mn atoms in different coordination environments. a) The DOS projected onto the $t_{2g}$ and $e_g$ orbitals of Mn atoms in octahedral coordination. b) The DOS projected onto the $t_{2g}$ and $e_g$ orbitals of Mn atoms in tetrahedral coordination. The values indicated in Figure S6 represent the integral of the DOS for the respective regions, corresponding to the number of electrons.

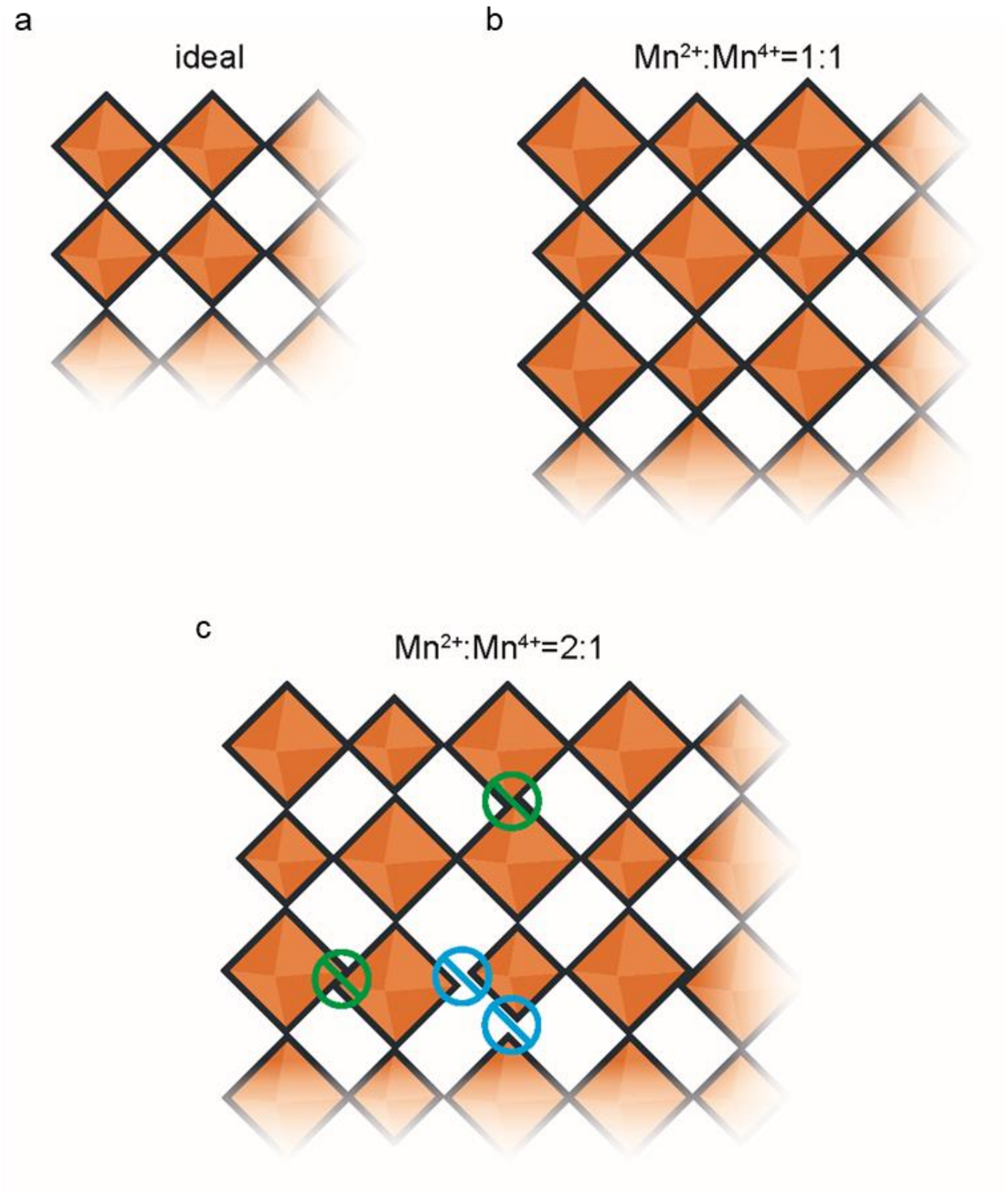

**Figure S7.** Schematics of octahedral networks under different configurations. a) The ideal oxygen corner-connected octahedral framework. b) The 1:1 population ratio of $d^5$ and $d^3$ configurations, forming a stable checkerboard-type network that maintains structural coherence. c) The 2:1 population ratio of $d^5$ and $d^3$ configurations, resulting in a disrupted octahedral network characterized by non-contacting vertices and and over-closed vertices.

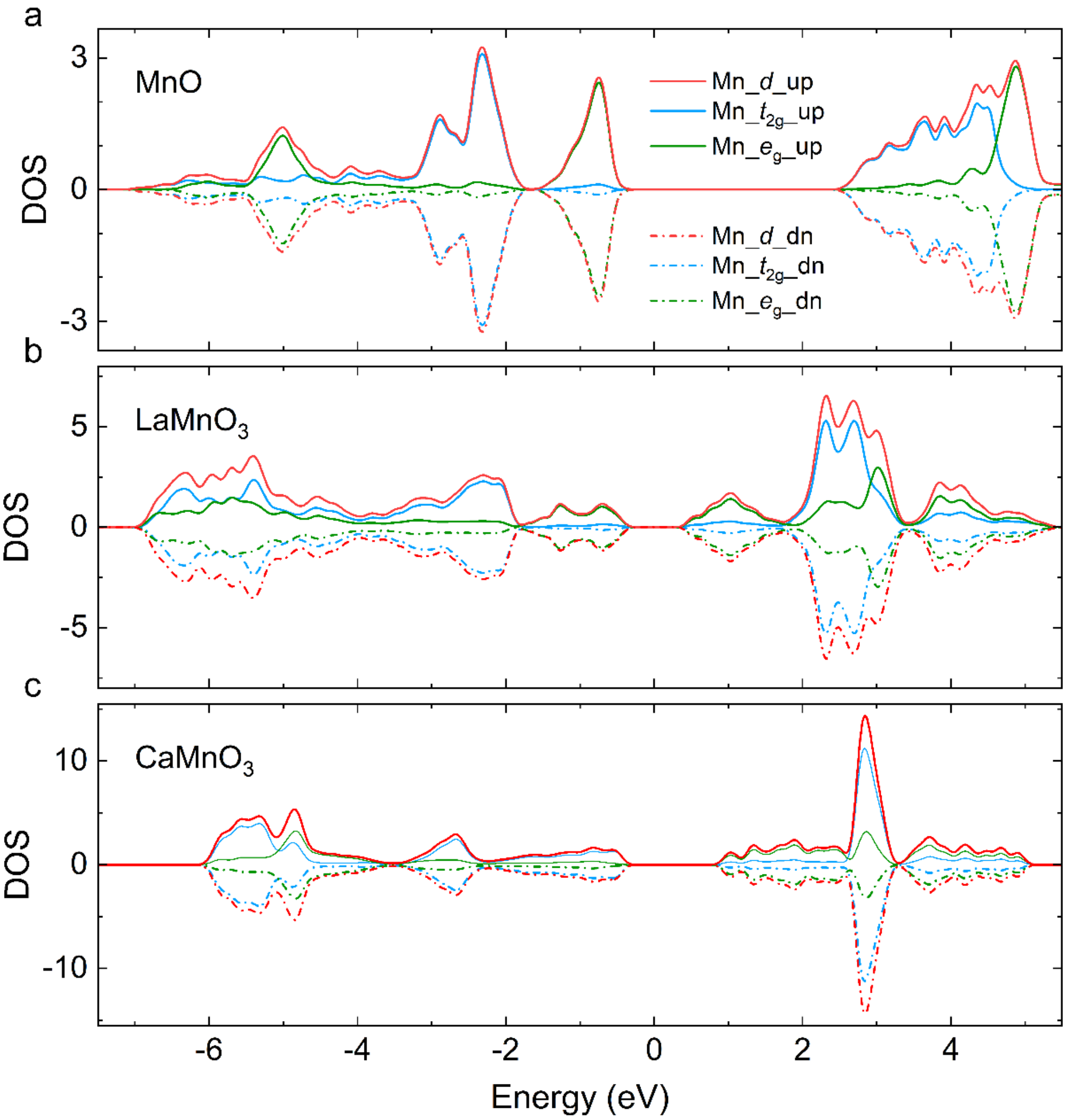

**Figure S8.** DFT calculations of single-valence perovskite manganites. DOS for a) MnO, b) $LaMnO_3$, and c) $CaMnO_3$. Solid lines represent the spin-up DOS, while dashed lines represent the spin-down DOS. The Mn 3*d* DOS are split into $t_{2g}$ and $e_g$ orbitals.

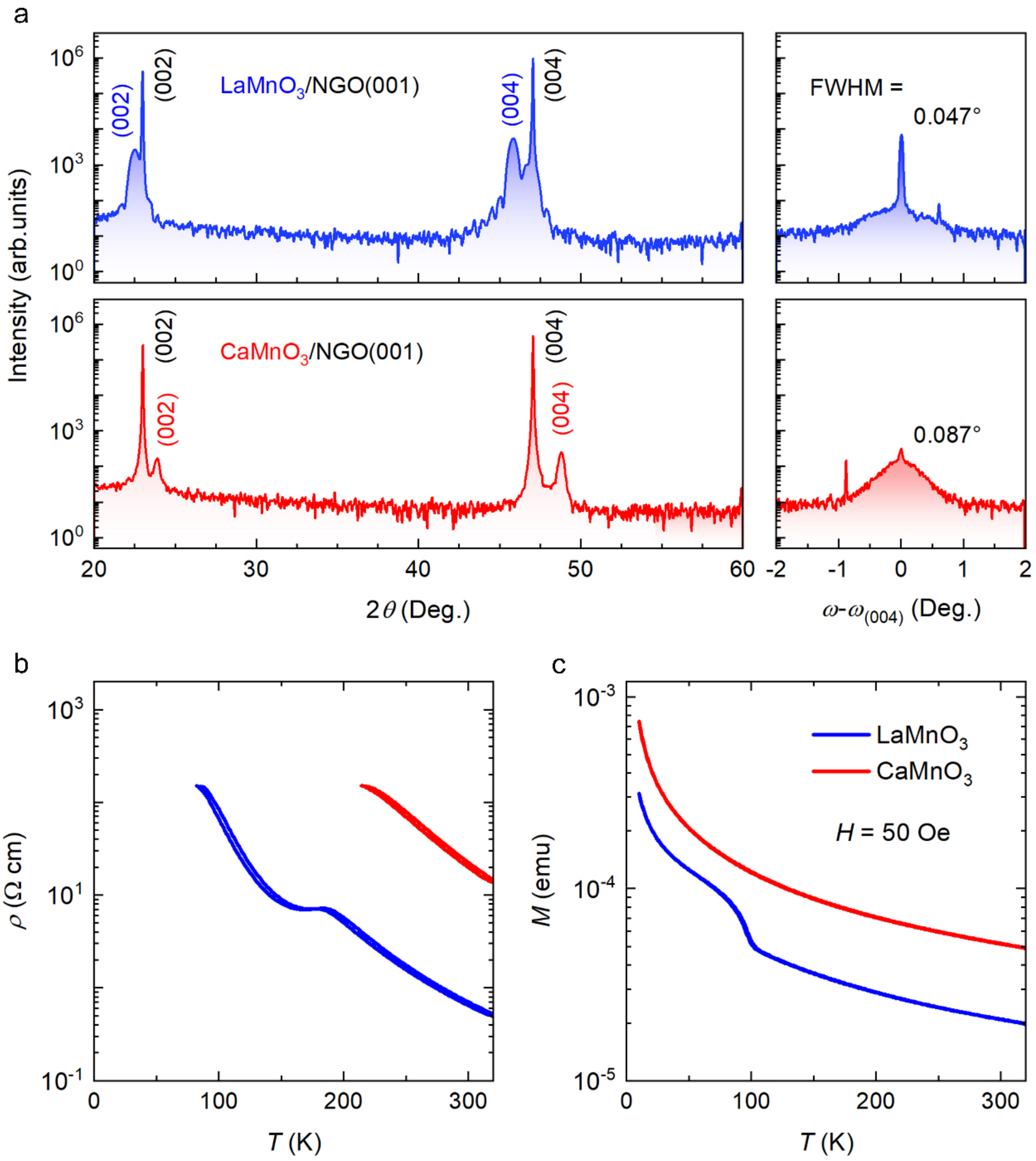

**Figure S9.** Structural characterizations and physical properties of single-valence perovskite manganites. a) XRD linear scans and rocking curves of $LaMnO_3$ and $CaMnO_3$ thin films grown on NGO(001) substrates. b) $\rho$-$T$ and $M$-$T$ curves of $LaMnO_3$/NGO(001) and $CaMnO_3$/NGO(001) thin films.

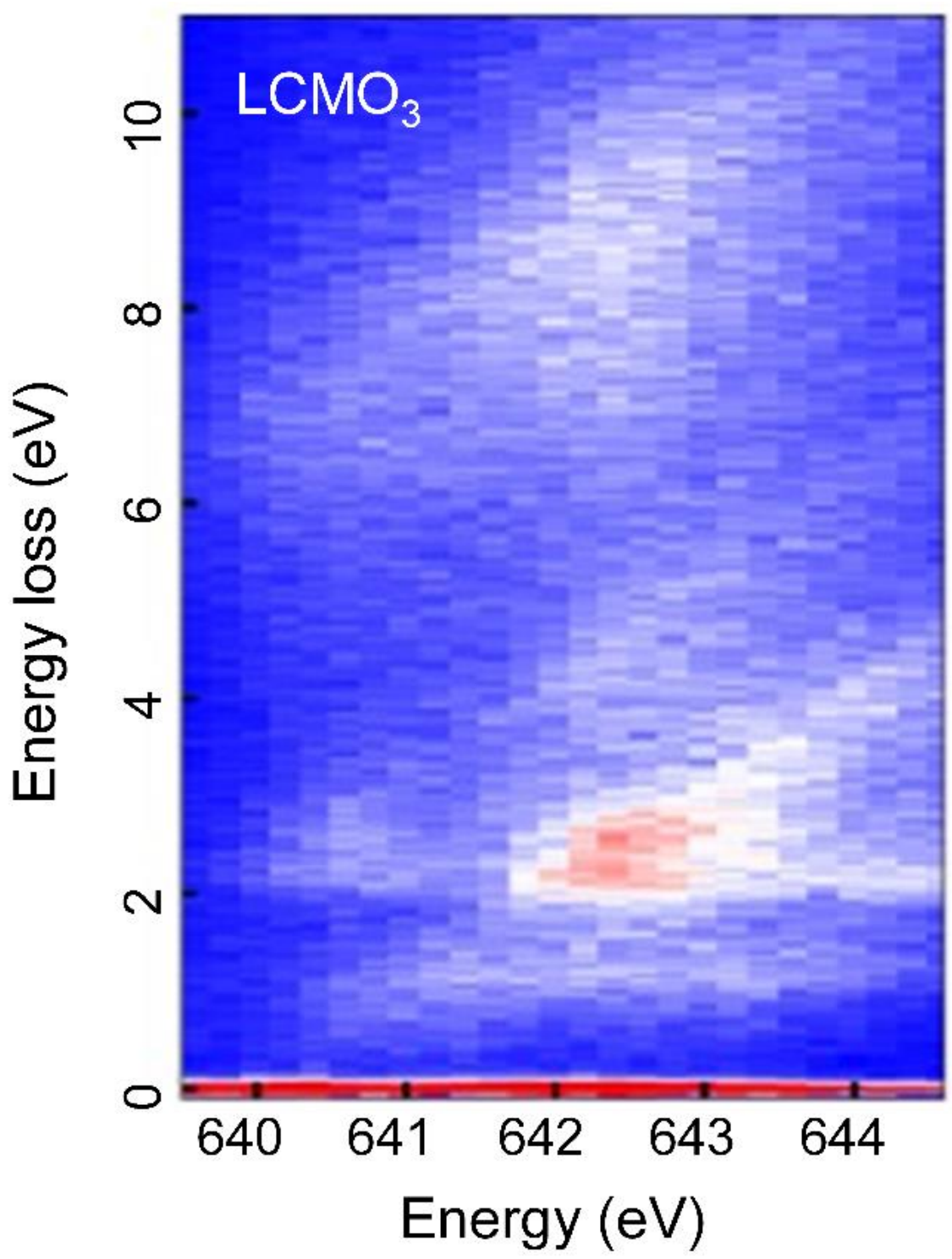

**Figure S10.** RIXS intensity color map of $LCMO_3$. RIXS intensity color map as a function of the incident x-ray energy and energy loss of $LCMO_3$. The measurement was taken at 40 K with the $\sigma$-polarization.

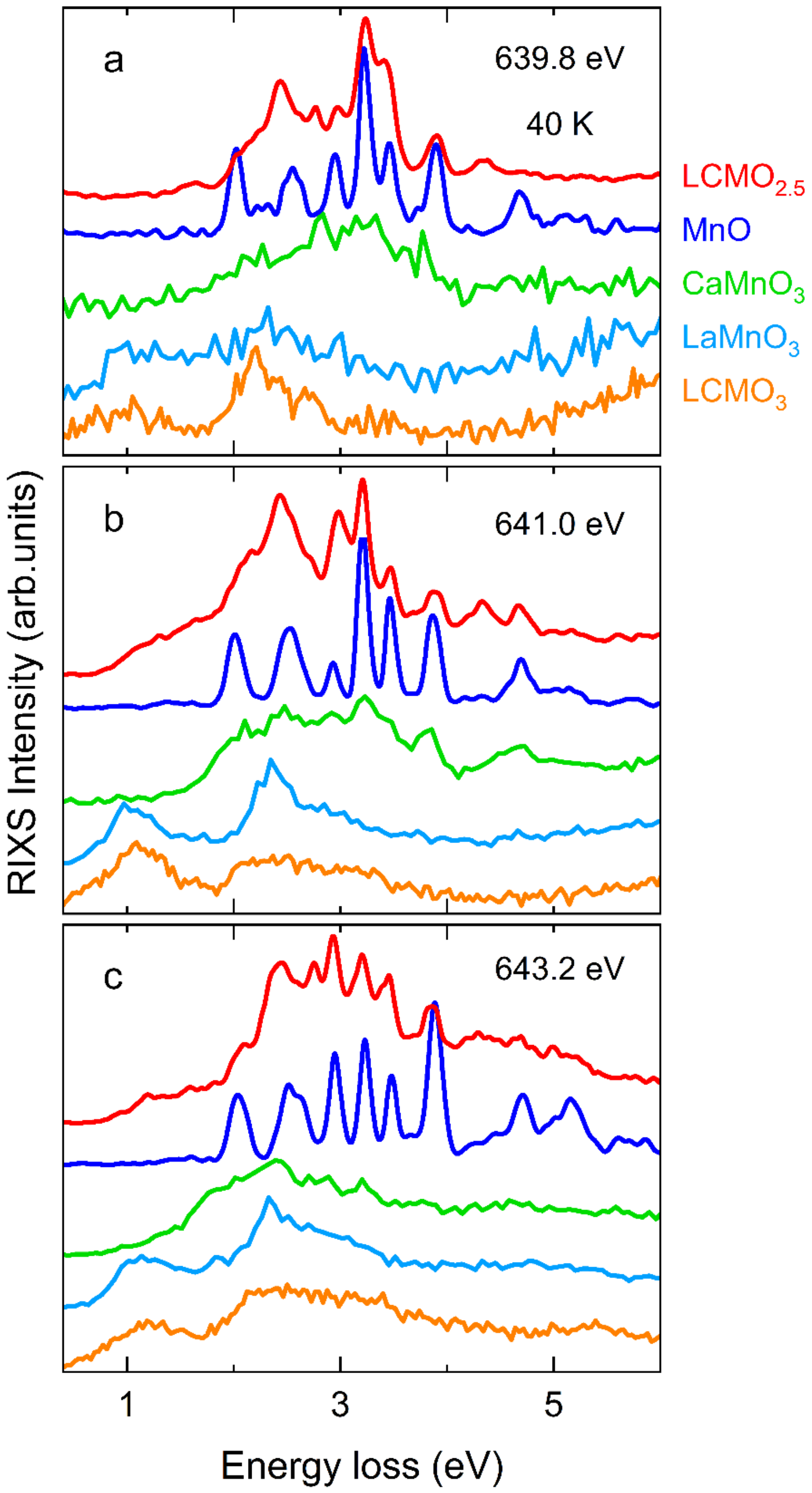

**Figure S11.** Fine scans of *dd* excitations. Comparison of fine RIXS spectra for $LCMO_{2.5}$, MnO, $LaMnO_3$, $CaMnO_3$, and $LCMO_3$ at selected incident energies of a) 639.8, b) 641.0 and c) 643.2 eV. All RIXS measurements are taken at 40 K.